\documentclass[%
reprint, superscriptaddress, showpacs,preprintnumbers, amsmath,amssymb, prl,
]{revtex4-1}

\usepackage[dvipdfmx]{graphicx}
\usepackage{dcolumn}
\usepackage{lineno}

\usepackage{bm}
\usepackage{color}
\usepackage{txfonts}
\usepackage{multibib}

\renewcommand{\figurename}{{\bf Fig.}}
\renewcommand\thefigure{{\bf \arabic{figure}}}

\newcommand{\vphi}{\varphi}

\newcommand{\be}{\begin{equation}}      
\newcommand{\ee}{\end{equation}}      
\newcommand{\bea}{\begin{eqnarray}}      
\newcommand{\eea}{\end{eqnarray}}

\newcites{Methods}{---}
\begin{document}

\title{\large  Chirality-driven edge flow and non-Hermitian topology in active nematic cells}

\author{Lisa Yamauchi$^{*}$}%
\affiliation{Nonequilibrium Physics of Living Matter RIKEN Hakubi Research Team, RIKEN Center for Biosystems Dynamics Research, 2-2-3 Minatojima-minamimachi, Chuo-ku, Kobe 650-0047, Japan.}%
\author{Tomoya Hayata$^{*}$}%
\affiliation{Theoretical Research Division, Nishina Center, RIKEN, Wako, Saitama 351-0198, Japan.}%
\affiliation{Department of Physics, Keio University, 4-1-1 Hiyoshi, Kouhoku–ku, Yokohama 223–8521, Japan.}%
\author{Masahito Uwamichi}
\affiliation{Department of Physics, The University of Tokyo, Bunkyo-ku, Tokyo 113-0033, Japan.}%
\author{Tomoki Ozawa$^\dag$}%
\affiliation{Advanced Institute for Materials Research, Tohoku University, Sendai 980-8577, Japan.}%
\affiliation{RIKEN Interdisciplinary Theoretical and Mathematical Sciences Program, 2-1 Hirosawa, Wako 351-0198, Japan.}%
\author{Kyogo Kawaguchi$^\dag$}%
\affiliation{Nonequilibrium Physics of Living Matter RIKEN Hakubi Research Team, RIKEN Center for Biosystems Dynamics Research, 2-2-3 Minatojima-minamimachi, Chuo-ku, Kobe 650-0047, Japan.}%
\affiliation{RIKEN Cluster for Pioneering Research, 2-2-3 Minatojima-minamimachi, Chuo-ku, Kobe 650-0047, Japan.}%
\affiliation{Universal Biology Institute, The University of Tokyo, Bunkyo-ku, Tokyo 113-0033, Japan.}%
\date{\today}

\begin{abstract}
Many of the biological phenomena involve collective dynamics driven by self-propelled motion and nonequilibrium force (i.e., activity) that result in features unexpected from equilibrium physics~\cite{marchetti_hydrodynamics_2013}. On the other hand, biological experiments utilizing molecular motors~\cite{schaller_polar_2010,sanchez_spontaneous_2012}, bacteria~\cite{zhang_collective_2010,nishiguchi2017long}, and mammalian cells~\cite{duclos_perfect_2014,saw_topological_2017,kawaguchi_topological_2017} have served as ideal setups to probe the effect of activity in materials and compare with theory~\cite{Gompper2020}. As has been established, however, biomolecules are chiral in nature, which can lead to the chiral patterning of cells~\cite{heacock_clockwise_1977,hagmann_pattern_1993,diluzio2005escherichia,tamada_autonomous_2010,wan_micropatterned_2011} and even to the left-right symmetry breaking in our body~\cite{gros_cell_2009,lebreton_molecular_2018,nonaka_randomization_1998,juan_myosin1d_2018}. The general mechanism of how the dynamics of bio-matters can couple with its own inherent chirality to produce macroscopic patterns is yet to be elucidated. Here we report that cultured neural progenitor cells (NPCs), which undergo self-propelled motion with nematic cell-to-cell interactions~\cite{kawaguchi_topological_2017}, exhibit large scale chiral patterns when flowing out from containers made by gel. Moreover, a robust chiral cell flow is produced along the boundary when the NPCs are cultured on substrates with edges. Perturbation by actomyosin inhibitors allowed control over the chirality, resulting in the switching of the direction of the chiral patterning and boundary flow. As predicted by a hydrodynamic theory analogous to the non-Hermitian Schr\"{o}dinger equation, we find an edge-localized unidirectional mode in the Fourier spectrum of the cell density, which corresponds to the topological Kelvin wave~\cite{delplace_topological_2017,shankar_topological_2017}. These results establish a novel mechanism of flow that emerges from a pool of bipolar cells, and demonstrate how topological concepts from condensed matter physics can naturally arise in chiral active systems and multi-cellular phenomena. 
\end{abstract}

\maketitle

The mechanism of how chiral motions at the level of components can give rise to robust macroscopic patterning has recently gained much interest in nonequilibrium physics~\cite{tsai2005chiral,furthauer2012active,han2020statistical,soni2019odd}.
In biological systems, the chirality of cytoskeletons and molecular motors are typically responsible for the chiral patterns observed at the cellular~\cite{heacock_clockwise_1977,hagmann_pattern_1993,diluzio2005escherichia,tamada_autonomous_2010} and multi-cellular levels~\cite{wan_micropatterned_2011}, with some shown to be directly causing the left-right asymmetry in our body plans~\cite{gros_cell_2009,lebreton_molecular_2018,nonaka_randomization_1998,juan_myosin1d_2018}. 

Recent works~\cite{shankar_topological_2017,souslov_topological_2017,dasbiswas_topological_2018,souslov_topological_2019,sone2019anomalous,nash_topological_2015,yang2020robust} have elucidated how chirality in classical systems can lead to situations similar to topological insulators, a phenomenon heavily studied in quantum systems~\cite{hasan2010colloquium}. In these settings, waves are observed at the boundary or the interface of the system in a predictable manner from the celebrated bulk-edge correspondence~\cite{hatsugai_chern_1993} while the net flow is suppressed in the bulk. An intriguing example is the Kelvin wave in geophysics, which explains the robust equatorial wave driven by the chirality of the earth rotation and has recently been found to have a topological origin~\cite{delplace_topological_2017}.

Whether such robust waves due to chirality and topology exists in biological systems is an interesting question. Indeed, spontaneous flow is found in a wide range of scales in biology, from the cortical flow~\cite{mayer2010anisotropies}, chiral flow in bacteria ~\cite{beppu2020edge}, up to the collective cell migration \textit{in vivo}~\cite{gros_cell_2009,mayor2016front}. Yet, to our knowledge, there is still no example of topological waves playing a role in the biological context.

Here we use neural progenitor cells (NPCs), an active nematic system~\cite{kawaguchi_topological_2017,doostmohammadi_active_2018,duclos_perfect_2014,kemkemer_elastic_2000,duclos_spontaneous_2018,blanch-mercader_turbulent_2018,saw_topological_2017}, as a model to investigate how chirality of biomaterials can couple with activity to generate unidirectional flow.
We first establish that the collective dynamics of NPCs are chiral by an assay involving cells flowing out from a container made by gel. Next, we observe how the NPCs under the confined geometry with circular and linear boundaries exhibit chiral edge currents. By numerical simulation and the analysis of the hydrodynamic theory, we find that the edge flow is a natural consequence of a chiral active nematic confined within a region with boundary. We further find that the fluctuation in the cell density at the edge shows a pattern of the topological Kelvin mode, resembling the situation in geophysics~\cite{delplace_topological_2017}. The results clarify the scenario of chiral cells producing topological edge flow, thus expanding the possibility of applying condensed matter concepts in understanding multi-cellular phenomena.

\vspace{3mm}

\noindent
{\bf Chiral patterning in neural progenitors}

\begin{figure}[!hbt]
 \begin{center}
  \includegraphics[width=89mm]{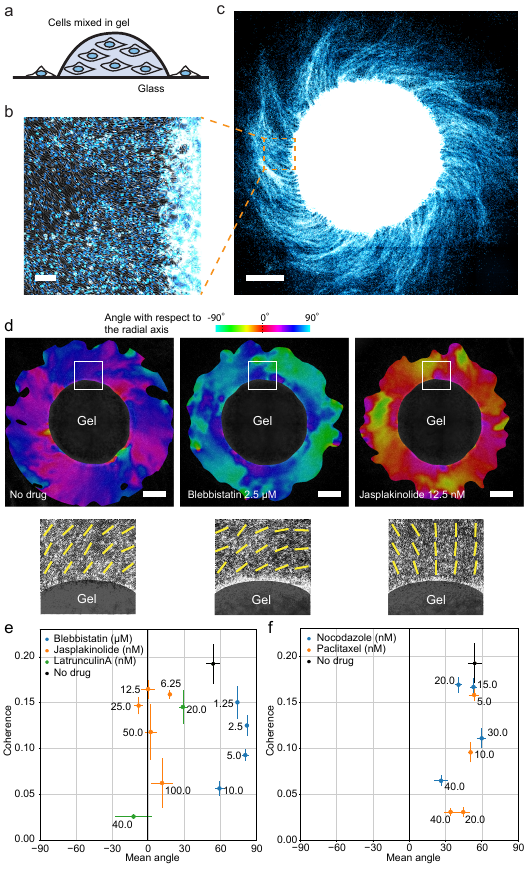}
 \end{center}
  \caption{\label{Fig1} {\bf Chiral patterning by  neural progenitor cells.}
{\bf a,} Schematic of experiment.
{\bf b,} Neural progenitor cells flowing out of the gel. Overlay of the phase contrast (grey) and H2B-mCherry (pseudocolour) channels. Scale bar: 100 $\mu$m.
{\bf c,} Example of chiral patterning. H2B-mCherry channel (pseudocolour). The white circular region is the gel, where the signal has saturated due to the high density of the cells. Scale bar: 1~mm.
{\bf d,} Top: drug-perturbed chirality quantified by image analysis. Colour code indicates the local angle of alignment with respect to the radial axis, with the centroid of the gel taken as the origin. Bottom: phase contrast image of the white boxed regions overlaid with the direction of local alignment. Scale bar: 1~mm.
{\bf e,} Chirality and coherence quantified for different dosages of actomyosin inhibitors. Coherence and mean angle are calculated using the tensor method (see Methods). Average of at least six gel positionss for each condition. Error bars: s.e.m.
{\bf f,} Same analysis for microtubule inhibitors.
}
\end{figure}

We first prepared a dense cell suspension in Matrigel, a commonly used extracellular matrix which is mainly composed of the laminin protein. When the gel droplets were deposited on glass and incubated in culture medium (Fig.~1a), cells spontaneously migrated out from the gel within a day. Because of the high cell density, the cells were nematically aligned soon after exiting from the gel (Fig.~1b).

After two days of culturing, when the cells had spread out from the gel due to migration and cell divisions, the cell population showed a clear chiral pattern (Fig.~1c). The angle of alignment, quantified by applying the tensor method~\cite{jahne_spatio-temporal_1993,rezakhaniha_experimental_2012} to the phase contrast images, became tilted at the positions further away from the center of the gel (Fig.~1d, left). The angle of tilting depended on the clone of the NPCs used in the experiment (Extended Data Fig.~1a), but the direction of chirality was always the same; the cell alignment pattern tilted toward the right respective to the radial axis.

We asked if perturbations on cytoskeleton dynamics can affect the chiral patterning, as has been tested in experiments using cell lines~\cite{wan_micropatterned_2011} and tissue dynamics in chicken development~\cite{gros_cell_2009}. The high concentration of cytoskeleton inhibitors prevented the spreading of the cells by blocking cell migration and division, and also disrupted the coherence of the nematic patterning (see Methods and Extended Data Fig.~1b). Nevertheless, several inhibitors, when used at low concentration, modulated the chirality of the pattern while maintaining the nematic ordering (Fig.~1d,e,f, Extended Data Fig.~1c). Of the tested inhibitors, Jasplakinolide, an actin stabilizer, and Latrunculin A, an actin polymerization inhibitor, weakened the right-handedness of the cell migration pattern~\cite{wan_micropatterned_2011}. On the other hand, Blebbistatin, a myosin II specific inhibitor, enhanced the right-handedness. Microtubule related inhibitors had less effect on chirality, and merely decreased the coherence of nematic order at higher concentrations. 

To see if the chirality exists at the single-cell level, we tracked the motion of NPCs at low cell density (Extended Data Fig.~1b). Although the motion of the cells is markedly different in the sparse condition, the cells showed a clear bias toward turning right ($\sim 0.1$ rad/hour, Extended Data Fig.~1d,e,f). However, we found no evidence of this single-cell-level chirality being affected by cytoskeleton inhibitors at least in the dosage range where cells were able to migrate.
This indicates that the inhibitors are affecting the chirality of cell-to-cell interactions, with possible mechanisms being the friction between rotating cell bodies as in the case of growth cone filopodia~\cite{tamada_autonomous_2010}, or the asymmetric sliding between neighboring cells~\cite{inaki2018chiral}.

\vspace{3mm}
\noindent
{\bf Chiral edge flow} 

\begin{figure}[!hbt]
 \begin{center}
  \includegraphics[width=89mm]{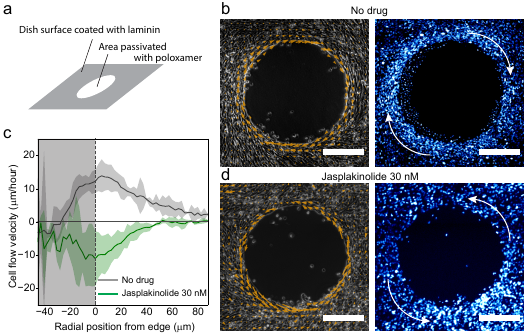}
 \end{center}
  \caption{\label{Fig2} {\bf Chiral edge flow of NPCs and its reversal.}
{\bf a,} Schematic of the dish surface coating experiment by laminin stamp and passivation.
{\bf b,} Phase contrast (left) and H2B-mCherry channel (right) images with the circular flow observed at the edges of open circles (500 $\mu$m diameter) of NPC culture. 
Orange arrows are proportional to the velocity of cell flow calculated by averaging the cell displacements within 30 $\mu$m square regions Scale bars: 200~$\mu$m.
{\bf c,} Average velocity in the azimuthal direction as a function of the radial coordinate. The origin of radial coordinate is taken at the edge of the stamp culture defined by the H2B-mCherry signal.
Average over 15 (no drug) and 16 (Jasplakinolide) stamp positions. Error bars: s.d. across three independent sets of experiments.
{\bf d,} Flow observed under the application of Jasplakinolide. Scale bars: 200~$\mu$m .
}
\end{figure}

We next asked if the chirality can be observed under other settings by employing a microcontact printing strategy to confine the cells~\cite{vedula_chapter_2014} (see Methods).
When we confined the cell migration area with a circular boundary by laminin coating and passivation (Fig.~2a), the cells showed unidirectional flow at the edge (Fig.~2b). By quantifying the average velocity by tracking the cells using the cell nucleus signal (H2B-mCherry), we found that the flow is strongly localized at the edge. This is remarkable since the individual cells do not have a net directionality; cells can migrate in either direction that respects the bipolar cell shape. Indeed, the cells flip their direction of motion stochastically once every 2-3 hours on average, which is why net flow is weak in the bulk except for the flow around topological defects~\cite{kawaguchi_topological_2017}. 
To confirm that this edge flow is arising from the same chiral effect in the macroscopic patterning observed in the gel experiment we perturbed the cells by Jasplakinolide. As expected, the edge flow switched its direction (Fig.~2c,d).

To further quantify the chiral edge flow in an even simpler setting, we again used the laminin stamps to contain the cells in stripes with various widths, $L$, which allowed the cells to show near-perfect nematic order  (Fig.~3a)~\cite{duclos_perfect_2014,kawaguchi_topological_2017,duclos_spontaneous_2018}.
Here, the cells showed similar edge flow as in the case of circular shaped boundaries; on the right edge, the cells were moving toward the positive $y$-direction, whereas on the left side the cells were migrating toward the negative (Fig.~3a).
A similar pattern of cell flow has been observed in retinal progenitor cells~\cite{duclos_spontaneous_2018}, although the chirality in NPCs was consistent and the edge flow was an order of magnitude faster, likely due to a different mechanism of chiral flow generation.
The cell density was higher near the boundary compared to the center region of the stripe (Fig.~3b, Extended Data Fig.~2a), and the cell flow in the $y$-direction was also concentrated at the edge (Fig.~3c).
Both of these features were observed even when cell division was blocked by Mitomycin C (Extended Data Fig.~2b), indicating that the flow and the inhomogeneous cell density is not due to the pattern of cell growth (Extended Data Fig.~2e).
Perturbation by Blebbistatin and Jasplakinolide resulted in slightly enhanced and reversed edge flow, respectively (Fig.~3d,e, Extended Data Fig.~2c,d), as expected from the gel-based assay (Fig.~1d).

\begin{figure*}[!hbt]
 \begin{center}
  \includegraphics[width=183mm]{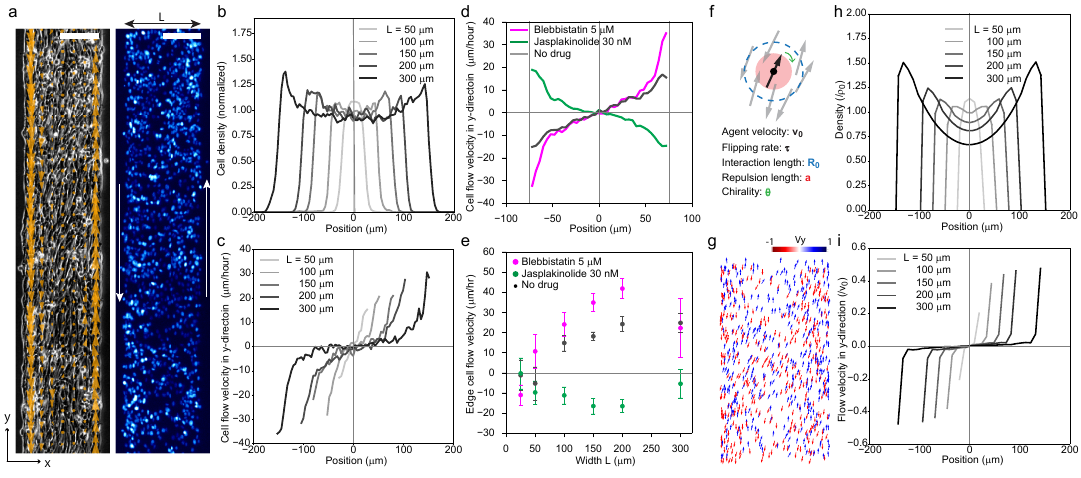}
 \end{center}
  \caption{\label{Fig3} {\bf Edge flow of NPCs in stripes.}
{\bf a,} Phase contrast (left) and H2B-mCherry channel (right) images of dense NPC culture confined in a stripe of width ($L$) 200 $\mu$m created by applying the laminin stamp on the dish substrate. Orange arrows are proportional to the velocity of cell flow calculated by averaging the cell displacements within 30 $\mu$m square regions. Scale bar: 100 $\mu$m.
{\bf b,c,} Cell density ({\bf b}) and mean velocity of cell flow in the $y$-direction ({\bf c}) as functions of the position $x$ for different stripe widths $L$. Cell density is divided by the mean density. Average over more than five stripes for each condition.
{\bf d,} Mean velocity of cell flow in the $y$-direction upon inhibitor applications. Average over more than five stripes, more than ten hours of tracking for each condition.
{\bf e,} Edge velocity for different stripe widths under inhibitor conditions. Error bars: s.e.m. across independent sets of experiments.
{\bf f,} Schematic of the numerical simulation of self-propelled rods with chirality.
{\bf g,} Snapshot of the numerical simulation for the condition mimicking the stripe boundary condition with $L=300$ $\mu$m.
{\bf h,i,} Density divided by the mean density $\rho_0$ ({\bf h}) and mean velocity of flow ({\bf i}) obtained by the numerical simulation. See Methods for the details of the parameters.
}
\end{figure*}

\vspace{3mm}
\noindent
{\bf Numerical simulation and theory of chiral active nematics}

To investigate the mechanism of the chiral edge flow, we conducted numerical simulation of an agent-based model that describes a dense active nematic system in two-dimensions.
In this model, the cells are represented by agents undergoing unidirectional motion with velocity $v_0$, stochastic flipping with average flipping time $\tau$, and rotational diffusion set by the constant $D_\phi$ (Fig.~3f). The agents interact with each other through repulsion and nematic alignment (Fig.~3g), with additional chirality controlled by the parameter $\theta$, which quantifies the rotation rate of the angle of the agents (Fig.~3f, see Methods for the detail of the numerical simulation).
The pattern within the stripe confinement, where the boundary condition was set to make the agents align to the edge, resembled the features observed in experiment when $\theta$ and $\tau$ were both nonzero (Fig.~3g, Extended Data Fig.~3a,b). The density profile, the pattern of alignment, and finite velocity near the edge were also reproduced when we set $\theta=0.2$ rad/hour. As expected, this value of chirality is the same order but larger than the chirality estimated from the cell tracks under low density (Extended Data Fig.~1f). 

From the experimental observations and numerical simulations, it is clear that the chiral symmetry breaking, $\theta \neq 0$, is key in the emergence of the edge flow.
To seek a theoretical explanation of this phenomenon, we turned to the hydrodynamic theory of dense active nematics, where the cell density, cell flow, and nematic order undergo time evolution simultaneously (see Supplementary Information).
Without chirality, the hydrodynamic equation has a steady-state solution describing a perfect nematic order, where the cell density is uniform ($\rho = \rho_0$), velocity field is zero ($ P_{x,y} =0$), and the nematic order is parallel to the edge ($Q_{yy} = -Q_{xx} = r >0$ and $Q_{xy} = 0$, where $Q$ is the nematic tensor and $r$ is positive parameter that quantifies the extent of alignment)~\cite{peshkov_boltzmann-ginzburg-landau_2014,peshkov_nonlinear_2012,patelli2019understanding}.
By adding small cell chirality, we obtain a linearized equation of time evolution, 
\begin{eqnarray}
\partial _t \Psi = \mathcal{M} \Psi + S.
\end{eqnarray}
Here, $\Psi = (\delta \rho/\rho_0 , \delta P_x, \delta P_y,  \delta Q_{xx},  \delta Q_{xy} )^T$ describes the fluctuations of density ($\delta \rho$), velocity field ($\delta P_{x,y}$) , and nematic tensor ($\delta Q_{xx}, \delta Q_{xy}$) around the perfect nematic ordered state. $\mathcal{M}$ is a $5 \times 5$ matrix including spatial derivatives, and $S = (0,0,0,0, 2 \theta r)$ is a constant global driving that arises due to $\theta$ (see Supplementary Information).
Importantly, $\mathcal{M}$ also includes $\theta$-dependent terms, which are analogous to the Coriolis force in geophysics~\cite{delplace_topological_2017} but with the origin being the chiral nature of the cell rather than the rotation of the earth.

Within the linear regime, the edge flow can be explained as the combination of chirality-induced deformation of nematic pattern and the pattern-induced active flow~\cite{simha2002hydrodynamic}.
First, the chirality of the cells induces a tilt in the steady-state nematic ordering (i.e., $\partial_x \delta Q_{xx,xy} \neq 0$).
Then, the spatial non-uniformity of the nematic order leads to the active force field, which is captured in the steady-state equation:
\begin{eqnarray}
\delta P_y  = - \eta \partial_x \delta Q_{xy}
\end{eqnarray}
where $\eta = {v_0}/{(4/\tau + 2D_\phi)}$ (see Supplementary Information).
The cell flow becomes significant only near the boundary since the tilt in the nematic order is localized at the edge.
The amplitude of the tilt in the nematic order, and therefore the edge flow, increases with the chirality of the cells.

The steady-state solution of equation (1) presents cell accumulation and flow that are qualitatively similar with the results of experiment and numerical simulation (Extended Data Fig.~4).
We note, however, that the situation of the experiment is expected to be out of the linear regime for large $L$ due to the relatively large chirality, $\theta$.
Nevertheless, we here study the linear equation expecting that the basic physics is preserved even when taking in the higher-order terms in the hydrodynamic equation (see Supplementary Information).

\begin{figure*}[!hbt]
 \begin{center}
  \includegraphics[width=183mm]{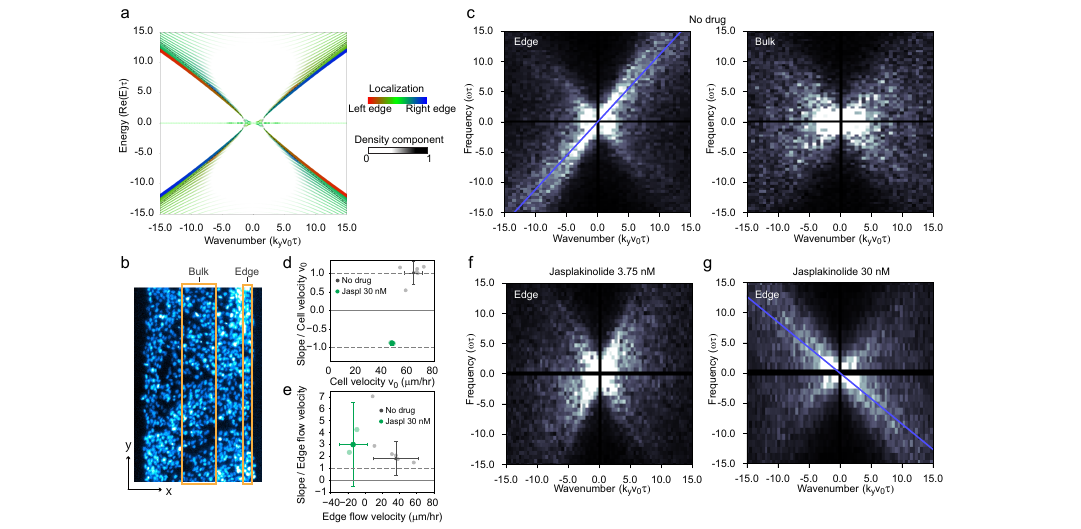}
 \end{center}
  \caption{\label{Fig4} {\bf Spectrum from the hydrodynamic theory and experiments.}
{\bf a,} Energy spectrum of the non-Hermitian Hamiltonian $\mathcal{H}$ calculated with the stripe boundary condition with $\theta=0.2$ rad/hour. Colour indicates the position of the localization of the modes, and the intensity of the colour indicates the contribution of the mode to the density fluctuation.
{\bf b,} Positions of the edge and bulk regions used for the calculation of the power spectrum. The image is showing the H2B-mCherry channel with $L=300$ $\mu$m. We defined the edge as the 25 $\mu$m wide region from the boundary and the bulk as the 100 $\mu$m wide region in the center.
{\bf c,} Power spectrum of the cell density calculated by the spatio-temporal Fourier transform of the H2B-mCherry signal.  Average over six regions. 
The blue line is the linear fit to the maxima of the power spectrum for each $k_y$. 
$v_0=72$ $\mu$m/hours was obtained from the data, and we set $\tau=2.5$ hours.
{\bf d,e,} Single cell velocity ({\bf d}) and the edge flow velocity ({\bf e}) versus the slope obtained as the fit to the power spectrum of the edge cell density. Each data point represents an independent experiment with $L=200$ $\mu$m or 300 $\mu$m that includes four or more regions. Error bars: s.d. across independent sets of experiments.
{\bf f,g,} Power spectrum upon application of Jasplakinlide with low ({\bf f,} 3.75 nM) and high ({\bf g,} 30 nM) concentrations. $v_0=50$ $\mu$m/hours ({\bf f}) and $v_0=47$ $\mu$m/hours ({\bf g}) were obtained from the data, and we set $\tau=2.5$ hours. Average over five ({\bf f}) and seven regions ({\bf g}).
}
\end{figure*}

\vspace{3mm}
\noindent
{\bf Spectrum of the linearized dynamics} 

Equation (1) can be considered as a non-Hermitian Schr\"{o}dinger equation with an external field.
We therefore calculated the complex spectrum of the non-Hermitian Hamiltonian, $\mathcal{H} = i\mathcal{M}$, under the stripe boundary condition similar to the experiment.
We found that the edge modes that are most significant in the cell density fluctuation has a pattern similar to the topological Kelvin mode obtained in fluid mechanics~\cite{delplace_topological_2017,shankar_topological_2017} (Fig.~4a).
By changing the extent of chirality, we further found that the amplitude of the edge localization becomes weaker as $\theta$ becomes close to zero, whereas the slope of the modes in the wavenumber-frequency plane does not change according to $\theta$ (Extended Data Fig.~5c,6c,7a).
This is also consistent with the edge-localized topological Kelvin mode, which we can identify through a simplified linear equation that assumes decoupling between the velocity and the nematic fields (see Supplementary Information). We obtain, as complex eigenvalues $E$ of $\mathcal{H}$,
\begin{eqnarray}
{\rm Re} E_\pm (k_y) = \pm \sqrt{\frac{1+r}{2}} v_0 \sqrt{ k_y^2 - k_0 ^2},
\end{eqnarray}
for $k_y^2 > k_0^2$ and zero otherwise, where $k_0 = [8 (1+r)]^{-1/2} \eta^{-1}$.
${\rm Re} E_+ (k_y)$ is plotted in Fig.~4a (dashed lines), which is in agreement with the spectrum of the full model.
Therefore, we expect that the mode of the density fluctuation at the edge should follow ${\rm Re} E_\pm (k_y) \simeq \pm \sqrt{(1+r)/2} v_0  k_y$ for large $k_y$, which is a $\theta$-independent dispersion relation.

To compare with this result, we calculated the power spectrum of the cell density by using the fluorescent signal of H2B-mCherry (Fig.~4b).
First, we found that there is a pattern of unidirectional mode observed only at the edges and not in the bulk region (Fig.~4c).
Second, the spectrum calculated from experiments with cytoskeleton perturbations showed that the group velocity of this unidirectional mode, obtained as the fitted slope to the spectrum for large $k_y$, is independent of the steady-state edge velocity and is rather correlated with the single cell velocity ($v_0$) of the NPCs under each condition (Fig.~4d,e).
Lastly, this unidirectional mode weakened (Fig.~4f) and switched direction (Fig.~4g) upon the application of Jasplakinolide.
These results are consistent with the interpretation that the unidirectional edge mode observed in the fluctuation of the cell density is the topological Kelvin wave.
We also found that the spectrum of the particle density fluctuation in the agent-based model follows a similar pattern (Extended Data Fig.~7b,c).

The main difference between our non-Hermitian setup and the previously studied Hermitian system~\cite{delplace_topological_2017} is that there is no band gap observed within the realistic range of parameters.
Nevertheless, we found that the band gap can be induced by assuming larger chirality in the same equation, where the bands can be found to have non-trivial topological Chern numbers (Extended Data Fig.~5, see Supplementary Information).
The mode described by Eq.~(3) existed irrespective of the energy gap, which matched with one of the topological edge modes in the case where the gap was open.
The gapped Hamiltonian can be continuously deformed to a Hermitian Hamiltonian without closing the gap, and the property of the edge localized mode did not change along this path.
The extent of edge localization of the unidirectional modes depended on $\theta$, and the regime of the experiment and simulation ($\theta=0.2$ rad/hours) is where there is significant edge localization of the modes without the energy gap.
Therefore, we find that the edge modes generated by the NPCs, the agent-based model simulation, and the non-Hermitian theory are all correspondents of the topological edge modes observed in Hermitian and gapped systems~\cite{delplace_topological_2017}.

In the band structure analysis of the non-Hermitian model, we further noticed that many of the modes identified in the bulk band for the periodic boundary condition are localized at the edge under the open boundary condition (Extended Data Fig.~5e). This is a phenomenon called the non-Hermitian skin effect, which has also recently been shown to have a topological origin~\cite{Yao2018,Okuma2020}. The direct consequence of this effect in chiral active matter is yet to be elucidated.

\vspace{3mm}
\noindent
{\bf Discussion and conclusion} 

In this work, we have shown that the chiral nature of the NPCs can produce not only a chiral spiral pattern in the collective cell dynamics but also a robust unidirectional steady flow and the Kelvin wave at the edge in a confined culture setting.
Results of numerical simulation and theory show that the dynamics at the edge, i.e., the geometry-induced steady-state active flow and the topological edge mode, are universal features of chiral active nematics.

Recent works have discovered that specific molecules such as myosins Ic, Id, and V are responsible for the chirality of cell motion~\cite{lebreton_molecular_2018,tamada_autonomous_2010,juan_myosin1d_2018}, which implies that the change in the balance between these molecules and less chiral myosins such as myosin II can lead to changes in the extent of chirality presented at the cell level.
Although the direction of cell chirality can be different across cell types and species, we expect to find similar edge effects in other real biosystems since the mechanism of flow and wave generation does not depend on the detail of the cell biology.
It will be particularly interesting to find {\it in vivo} examples of topological modes, such as surface flow inside cells and cell flow in tubes and organs.

\newpage
\ 
\newpage

\noindent
{\bf {\large Methods}}
\vspace{3mm}

\noindent
{\bf Neural progenitor cell culture}

\noindent
We used cell lines of NPCs, which were previously established from mice of the standard strain E14 ICR39,40. The cells were cultured as described previously~\cite{kawaguchi_topological_2017} with a few modifications. For normal culturing, we used DMEM/F12 without phenol red with HEPES (Thermo Fisher Scientific Inc. 11039021) and added N-2 MAX media supplement (100x, R\&D Systems, Inc. AR009), bFGF (20 ng/ml, FUJIFILM Wako Pure Chemical Corporation 060-04543), and EGF (20 ng/ml, Thermo Fisher Scientific Inc. 53003018). For coating of the dishes, we prepared a cold stock solution of 10 \% Matrigel (Corning 356231) in PBS (FUJIFILM Wako Pure Chemical Corporation 049-29793), mixed 30 $\mu$l of that solution per 2 ml of culture medium (final 0.15 \% Matrigel), poured onto plastic or glass-based dishes and pre-incubated in 37 deg. Cells were directly passaged onto the prepared dishes with desired cell density. Cells were frozen and stored in Stem Cell Banker (GMP grade, TAKARA BIO INC. CB045) at -150 deg.

\vspace{3mm}
\noindent
{\bf Inhibitor preparation}

\noindent
For cytoskeleton perturbation experiments, we used Cytochalasin D (FUJIFILM Wako Pure Chemical Corporation 037-17561, stock solution 1 mg/mL in DMSO), Nocodazole (Sigma-Aldrich Co. LLC M1404, stock solution 10 mM in DMSO), Paclitaxel (FUJIFILM Wako Pure Chemical Corporation 167-28161, stock solution 3 mM in DMSO), (-)-Blebbistatin (FUJIFILM Wako Pure Chemical Corporation 021-17041, stock solution 10 mM in DMSO), and Jasplakinolide (Toronto Research Chemicals Inc J210700, stock solution 1 mM in DMSO). We also used the cell cycle inhibitor Mitomycin C (1 mg/ml solution, Nacalai Tesque 20898-21).

\vspace{3mm}
\noindent
{\bf Gel drop assay}

\noindent
Cells were detached from dishes and spun down to make a pellet of $2.5 \times 10^6$ cells. The pellet was suspended in 50 $\mu$l ice-cold Matrigel which was stored in aliquots at -20 deg.
The cell suspension in gel was plated onto glass base dishes using pipette tips which were stored in the refrigerator until use. Each drop consisted of 2 $\mu$l of suspension.
After 3 minutes of incubation in 37 deg, the plate was filled with warm culture medium with additional Matrigel suspension (0.15\%) and left overnight in the incubator.
The next day, the media in the glass base dishes were changed to fresh media with or without the inhibitors.
After two to three days of culturing, the glass base dishes with the gel drops were placed under the fluorescent microscope (Zeiss AxioObserver7). Fluorescence and phase contrast images typically of size 7.3 mm x 7.3 mm were obtained by the 10x lens using the tiling acquisition setup.

\vspace{3mm}
\noindent
{\bf Image analysis of the chiral patterning}

\noindent
Colour maps representing the tilt angle of the cell alignment were obtained by the tensor method~\cite{kawaguchi_topological_2017,jahne_spatio-temporal_1993,rezakhaniha_experimental_2012}. For an acquired phase contrast image $I({\bf r})$, where ${\bf r}= (x,y)$ denotes the two-dimensional position (in pixels), we calculated the differential tensor:
\begin{eqnarray}
J({\bf r},t) :=  \left( \begin{array}{cc}
(\Delta_x I)^2 & \Delta_x I\Delta_y I  \\
\Delta_x I\Delta_y I & (\Delta_y I)^2 \end{array} \right).
\end{eqnarray}
with
\begin{eqnarray}
\Delta_x I  &:=& I(x+1,y)-I(x-1,y) \\
\Delta_y I &:=& I(x,y+1)-I(x,y-1).
\end{eqnarray}
After applying a spatial Gaussian filter to $J$ to obtain
\begin{eqnarray}
\widetilde{J}({\bf r},t) :=  \left( \begin{array}{cc}
\widetilde{J}_{xx}  & \widetilde{J}_{xy}   \\
\widetilde{J}_{xy}  & \widetilde{J}_{yy}  \end{array} \right),
\end{eqnarray}
where the size of the filter was 65 $\mu$m, we calculated the local principal angle of alignment
\begin{eqnarray}
\theta ({\bf r}) &:=& \frac{1}{2}   \arctan \left( \frac{ 2 \widetilde{J}_{xy}}{\widetilde{J}_{xx}-\widetilde{J}_{yy}} \right ),
\end{eqnarray}
and the coherence
\begin{eqnarray}
C({\bf r}) &:=& \frac{(\widetilde{J}_{xx}-\widetilde{J}_{yy})^2+4 {\widetilde{J}}_{xy}}{(\widetilde{J}_{xx}+\widetilde{J}_{yy})^2}.
\end{eqnarray}
The coherence, which takes values between 0 and 1, quantifies the extent of anisotropy in the image. For the phase contrast images of the NPCs, the perfectly aligned situation corresponded to $C \sim 0.3$, as shown in Extended Data Fig.~1b.

For Fig.~1d, we masked out the regions of the gel and the regions where cells are absent by setting upper and lower thresholds in the intensity of the H2B-mCherry channel.
We calculated the mean coherence and tilt angle by averaging the gradient tensor over all the pixels in each image using the same mask (Fig.~1d,e).
All analyses were conducted using a custom Python code.

\vspace{3mm}
\noindent
{\bf Low cell density experiment}

\noindent
NPCs tend to become unhealthy under low cell density conditions, which can be rescued by applying conditioned media.
We prepared conditioned media by taking the supernatants from the plates with sub-confluent cell density. Typically, we plated NPCs at a quarter of confluent density, changed medium the next day, and collected the supernatant the day after. The supernatant was spun down 800 g 5 minutes, filtered 0.22 $\mu$m, and aliquoted and frozen down in -20 deg.

For low cell density experiments, we plated $2\times 10^4$ cells per glass base dish in 1.5 ml thawed and warmed conditioned media with 0.15\% Matrigel and waited three hours until the cells adhered to the dish surface.
We added 0.5 ml of conditioned media with or without cytoskeleton inhibitors and left the dish overnight in incubation.
The next day, the glass base dishes with low cell density were put on the confocal microscope (LSM800, Zeiss) to take tiled images of the fluorescent channel for H2B-mCherry every 5 minutes using a 10x lens.
For the tracking of cells, we used the TrackMate plugin included in the Fiji package~\cite{tinevez_trackmate:_2017}.

From the obtained single-cell tracks, we calculated the average angle change as a function of time (Extended Data Fig.~1f) using a custom Python code. Using the angle of displacement $\theta_i (t)$ of the $i$-th cell at between time frames $t$ and $t+1$, we calculated the average angle change by
\begin{eqnarray}
\Delta \bar{\theta} (t) &:=& \arctan \left( \frac{\langle \sin (\theta_i(s+t) - \theta_i(s)) \rangle}{\langle \cos  (\theta_i(s+t) - \theta_j(s)) \rangle} \right ),
\end{eqnarray}
where the average $\langle ... \rangle$ is taken over the different cells ($i$) and time ($s$).

\vspace{3mm}
\noindent
{\bf Microcontact printing experiment}

\noindent
For the micropatterning experiment~\cite{vedula_chapter_2014}, we first designed photomasks and ordered printing on positive film sheets (Tokyo Lithmatic Corporation, Tokyo).
Next, we prepared a substrate of negative photoresist (SU-8 3025, MicroChem) on silicon wafer by spin coating, applied the photomask after soft baking, and exposed to UV light using an LED spotlight.
After developing and baking the photoresist pattern by the standard procedure, we made stamps by pouring polydimethylsiloxane (PDMS, SYLGARD 184 Silicone Elastomer Kit, Dow Corning 98-0898) onto the pattern and incubating at 80 deg for more than two hours.

To make dishes with micropatterned laminin, we first cleaned the PDMS stamp surface using an oxygen plasma cleaner for two minutes, deposited 50 $\mu$l of cold 20\% Matrigel solution (in PBS) on its surface, and incubated 30 mins at room temperature.
We then removed the matrigel solution by an aspirator, and gently placed the stamp on the dish and gently pushed with a finger to remove air bubbles before applying a 5-6 g weight.
We used $\mu$-Dish (35 mm High, uncoated, Nippon Genetics ib81151, ibidi 81151) instead of glass base dishes since they had better passivation properties.
After carefully removing the stamp, we applied 2 ml of 0.2 \% Pluronic F-127 solution (Molecular Probes P6866) to the entire dish and incubated 60 mins at room temperature.
We then washed the dish four times with 1 ml PBS, and lastly poured in the cell solution at 1/4 of confluency and left overnight in the incubator.

The next day, we washed the dish three times using culture media to remove the floating cells, and applied 2 ml of conditioned media with or without the inhibitors before the live imaging.

\vspace{3mm}
\noindent
{\bf Calculation of cell flow and density}

\noindent
To calculate cell flow, we first generated the tracks of the individual cells by using the H2B-mCherry signal and the  TrackMate plugin included in the Fiji package~\cite{tinevez_trackmate:_2017}.
We set a threshold in the detection of cell nuclei so that only the brightest 10-30 \% of the cells in the culture were tracked, in order to reduce the probability of mis-tracking.
Using the data of tracks, we calculated the local average velocity of flow by averaging in the azimuthal direction (Fig.~2c), $y$-direction (Fig.~3c, Extended Data Fig.~2),  or within the boxes of size 30 $\mu$m (Figs.~2b,d,3a).
We calculated the single cell velocity ($v_0$) for each condition from the mean-square displacement obtained from the tracks.
For the cell density, we used either the counts of the cells in the tracks (Fig.~3b, Extended Data Fig.~2a,b,c,d) or the sum of the H2B-mCherry signal by subtracting the background (Extended Data Fig.~2e).
All analyses were conducted using a custom Python code.

\vspace{3mm}
\noindent
{\bf Numerical simulation of the agent-based model}

\noindent
In the numerical simulations (Figs.~3f,g,h,i, Extended Data Fig.~7b,c), we used a model where the cells are represented by agents~\cite{chate_simple_2006}; the $i$-th agent at time $t$ is characterized by the two-dimensional coordinates $\bm r_i(t)=(x_{i}(t),y_{i}(t))$ and the direction of motion (director) $\bm p_i(t)=(p_{ix}(t),p_{iy}(t))$. Here, $\bm p_i(t)$ is a two-dimensional unit vector, $|\bm p_i(t)|=1$.
The time-evolution of $\bm r_i(t)$ and $\bm p_i(t)$ is defined as
\bea
\bm r_i(t+\Delta t) &=& \bm r_i(t)+v_0 \Delta t \bm p_i(t) ,
\\
 \tilde{\bm p}_i(t) &=&\bar{\bm p}_i(t)+\sum_{j\neq i}\bm F_{ji}(t) ,
\\
\bm p_i(t+\Delta t) &=&R_{\eta,\tau}\left[ \tilde{\bm p}_i(t)/\left| \tilde{\bm p}_i(t)\right|\right] ,
\eea
where $v_0$, $\bar{\bm p}_i$, and $\bm F_{ji}$ are the velocity of agents, mean director in the neighbored of the $i$-th agent, and exclusion force from the $j$-th agent to the $i-$th agent, respectively.
$R_{\eta,\tau}[\bm O]$ represents the random noise and velocity flipping.
By operating $R_{\eta,\tau}$, a unit vector $\bm O=(\cos\vphi,\sin\vphi)$ is mapped to $R_{\eta,\tau}[\bm O]=(\cos(\vphi+\eta_1\pi+\eta_2\pi),\sin(\vphi+\eta_1\pi+\eta_2\pi))$, where $\eta_1\in[-h_0,h_0]$ is a uniform random noise, and $h_0(\leq1)$ is a parameter characterizing the randomness of the direction.
$\eta_2$ is another stochastic variable which takes 0 or 1, with the probability of taking 1 given by $\Delta t/\tau$, where $\tau$ is the velocity flipping rate.

The mean director (molecular field) is given by
\bea
\bar{\bm p}_i(t)&=&\sum_{j}n_{ij}\left[\left(\bm p_i\cdot\bm p_j\right)+\theta\Delta t\left(\bm p_i\times\bm p_j\right)_z\right]\bm p_j \notag\\
&& -\sum_{j}\theta\Delta t n_{ij}\left(\bm p_i\cdot\bm p_j\right)\hat{z}\times\bm p_j, 
\label{eq:hamiltonian}
\eea
where $(\bm p_i\times\bm p_j )_z=p_{ix}p_{jy}-p_{iy}p_{jx}$, $\hat{z}\times\bm p_j= (-p_{jy},p_{jx})$
and
\be
n_{ij}=
\begin{cases}
1 \;\;\text{if}\;\;|\bm x_i-\bm x_j|<R_0\\
0 \;\;\text{if}\;\;|\bm x_i-\bm x_j|>R_0
\end{cases} .
\ee
Note that the summation includes the $i$-th agent.
The first term in Eq.~\eqref{eq:hamiltonian} represents the nematic alignment, while the second and third terms represent the chirality-driven alignments parameterized by $\theta$.
The chirality-driven alignment is introduced so that the $i$-th agent nematically interacts with the $j$-th agent with a tilt. 
To derive the second and third terms in Eq.~\eqref{eq:hamiltonian}, we consider the rotation of the $j$-th agent around the $z$ axis ($\bm p_j\rightarrow  \bm p_j^\prime= \bm p_j-\theta \Delta t \hat{z}\times \bm p_j$), and plug $\bm p_j^\prime$ into the nematic alignment term.
The autonomous rotation of agents is included in the last term in Eq.~\eqref{eq:hamiltonian}.

In addition to the chiral nematic alignment, we included the exclusion force $\bm F_{ji}$ to avoid over-clustering of the alignments.
We assume the short-ranged repulsive interaction:
\be
\bm F_{ji}=k\Theta\left(a-|\bm x_{ij}|\right)\left(a-|\bm x_{ij}|\right)\bm x_{ij}/|\bm x_{ij}| ,
\ee
where $\bm x_{ij}=\bm x_i-\bm x_j$, and $\Theta(x)$ is the Heaviside step function.
$k$ and $a$ are the strength and the length scale of the repulsive force, respectively.

We assume that the system is finite along the $x$-direction  ($x\in[-L/2,L/2]$), and is periodic in the $y$-direction. 
Near the edges, the cells align their shape with the boundary and can only move parallel to the edge. To mimic this boundary effect, we impose the following condition for the cells at position $-L/2 < x_{ix}<-L/2+0.50a$ or $L/2-0.50a <x_{ix}<L/2$:
\be
\tilde{\bm p}_i(t)=
{\rm sign}(p_{iy})\hat{y}.
\label{eq:edge}
\ee
Here, $a$, and $L$ are the size of the agent (same as the length scale of the repulsive force), and length of the system along the $x$-direction, respectively.

The numerical conditions we used for Figs.~3h,i,4h, Extended Data Figs.~3a,b,7b,c are
$\rho_0=4000$ cells/mm$^2$, $v_0=$40 $\mu$m/hour, $\Delta t=$0.02 hours, $h_0=$0.06, $R_0=$35 $\mu$m, $k=$0.12 $\mu{\rm m}^{-1}$, $a=$12.5 $\mu$m, with $\tau$=2.5 hours or $\infty$ (no flipping), $\theta$=0.2 rad/hour or 0 rad/hour (zero chirality).
We used a custom C++ code to run the simulation, and made the plots using a custom Python code.

\vspace{3mm}
\noindent
{\bf Spectral analysis of image data and simulation}

\noindent
To obtain the power spectrum of the cell density, we used the timelapse image from the H2B-mCherry channel $\rho({\bm r},t)$, where ${\bm r} = (x,y)$ and $t$ denote the two-dimensional position (in pixels) and time (frame number), respectively.
We first took the stripe experiment data with $L=300$ $\mu$m, and cropped out the 25 $\mu$m edge regions from the left and right, and the 100 $\mu$m width region from the center.
The signal was normalized by subtracting the mean to minimize the spatio-temporal inhomogeneity in the images: $\tilde{\rho}({\bm r},t) = \rho'({\bm r},t) -  \sum_t \rho'({\bm r},t) / T$ with $\rho'({\bm r},t) = \rho({\bm r},t) - \sum_{\bm r} \rho({\bm r},t) / N_A $, where $T$ and $N_A$ are the number of time frames and the number of pixels in the sums, respectively.
After taking the sum over $x$, $\tilde{\rho}(y,t) = \sum_x \tilde{\rho}({\bm r},t)$, we calculated the discrete Fourier transform of $\tilde{\rho}(y,t)$ to obtain the power spectrum $P(k_y,\omega)$, and averaged them across multiple regions for each conditions (Figs.~4c,f,g).
To obtain the power spectrum of the edge, we assumed 180$^\circ$ rotational symmetry and flipped the $y$-axis for the left edge spectrum and added it to the spectrum of the right edge, before the averaging over regions to generate the plots in Fig.~4c,f,g.

To obtain the group velocity from the power spectrum, we first obtained $\Omega(k_y) = {\rm Argmax}_\omega P(k_y,\omega)$, and then fit $\Omega(k_y)$ with a linear function, $\Omega(k_y) = v_s k_y$ at $k_y v_0 \tau > 5$ to obtain the slope $v_s$.

For the Fourier analysis of the simulation, we took data corresponding to the $L=$300 $\mu$m setup, and first calculated the density of the particles along $y$ at each time points at the edge (25 $\mu$m wide) and the bulk (100 $\mu$m wide) regions (Fig.~4b) with the binning size of 25 $\mu$m in the $y$-direction.
We calculated the power spectrum of these one-dimensional densities by Fourier transform to obtain Extended Data Fig.~7b,c.
All analyses were conducted using a custom Python code.

\vspace{3mm}
\noindent
{\bf Acknowledgements} 

We thank Allon M. Klein for the support in the original experiments, and Masaki Sano, Tetsuya Hiraiwa, Takahiro Sagawa, Kazuki Sone, and Daiki Nishiguchi for the scientific discussions, and Kyosuke Adachi, Takaki Yamamoto, and Yosuke Fukai for commenting on the manuscript. T.O. is supported by JSPS KAKENHI Grant Number JP20H01845, JST PRESTO Grant Number JPMJPR19L2, JST CREST Grant Number JPMJCR19T1, and the Interdisciplinary Theoretical and Mathematical Sciences Program (iTHEMS) at RIKEN. K.K is supported by JSPS KAKENHI Grants No. JP18H04760, No. JP18K13515, No. JP19H05275, No. JP19H05795, and the Human Frontier Science Program. The numerical calculations have been performed on cluster computers at RIKEN iTHEMS.

\vspace{3mm}
\noindent
{\bf Author contributions}

L.Y. and K.K. conducted the experiments and analyzed the data. M.U, developed the micropatterning experiment. T.H. performed numerical simulations of the agent-based model, and T.O. calculated the band structure of the hydrodynamic model. T.H., T.O., and K.K. developed the theory and wrote the supporting information. K.K. wrote the manuscript with input from all authors.

\newpage
\ 
\newpage

\newpage

\onecolumngrid

\makeatletter
\providecommand \@ifxundefined [1]{%
 \@ifx{#1\undefined}
}%
\providecommand \@ifnum [1]{%
 \ifnum #1\expandafter \@firstoftwo
 \else \expandafter \@secondoftwo
 \fi
}%
\providecommand \@ifx [1]{%
 \ifx #1\expandafter \@firstoftwo
 \else \expandafter \@secondoftwo
 \fi
}%
\providecommand \natexlab [1]{#1}%
\providecommand \enquote  [1]{``#1''}%
\providecommand \bibnamefont  [1]{#1}%
\providecommand \bibfnamefont [1]{#1}%
\providecommand \citenamefont [1]{#1}%
\providecommand \href@noop [0]{\@secondoftwo}%
\providecommand \href [0]{\begingroup \@sanitize@url \@href}%
\providecommand \@href[1]{\@@startlink{#1}\@@href}%
\providecommand \@@href[1]{\endgroup#1\@@endlink}%
\providecommand \@sanitize@url [0]{\catcode `\\12\catcode `\$12\catcode
  `\&12\catcode `\#12\catcode `\^12\catcode `\_12\catcode `\%12\relax}%
\providecommand \@@startlink[1]{}%
\providecommand \@@endlink[0]{}%
\providecommand \url  [0]{\begingroup\@sanitize@url \@url }%
\providecommand \@url [1]{\endgroup\@href {#1}{\urlprefix }}%
\providecommand \urlprefix  [0]{URL }%
\providecommand \Eprint [0]{\href }%
\providecommand \doibase [0]{http://dx.doi.org/}%
\providecommand \selectlanguage [0]{\@gobble}%
\providecommand \bibinfo  [0]{\@secondoftwo}%
\providecommand \bibfield  [0]{\@secondoftwo}%
\providecommand \translation [1]{[#1]}%
\providecommand \BibitemOpen [0]{}%
\providecommand \bibitemStop [0]{}%
\providecommand \bibitemNoStop [0]{.\EOS\space}%
\providecommand \EOS [0]{\spacefactor3000\relax}%
\providecommand \BibitemShut  [1]{\csname bibitem#1\endcsname}%
\let\auto@bib@innerbib\@empty

\renewcommand{\theequation}{S\arabic{equation}}
\setcounter{equation}{0}

\begin {center} 
	\textbf{ \large Supplementary Information for ``Chirality-driven edge flow and non-Hermitian topology in active nematic cells"}\\ [.1cm]
	
	{Lisa Yamauchi$^*$, Tomoya Hayata$^*$, Masahito Uwamichi, Tomoki Ozawa$^\dag$, and Kyogo Kawaguchi$^\dag$}\\ [.1cm]
	{(Dated: \today)} \\
\end {center}

\date{\today}

\section{Hydrodynamic model and its topology}

We derive the hydrodynamic model of the chiral active nematic system using the Boltzmann-Ginzburg-Landau approach~\cite{peshkov_boltzmann-ginzburg-landau_2014,peshkov_nonlinear_2012,patelli2019understanding}. 
We start from the Boltzmann equation
\begin{align}
	\partial_t f(\mathbf{r},\phi,t) + v_0 \mathbf{e}_\phi \cdot \nabla f(\mathbf{r},\phi,t) - \theta \partial_\phi f(\mathbf{r},\phi,t)
	=
	\frac{1}{\tau}\left[ f(\mathbf{r},\phi+\pi, t) - f(\mathbf{r},\phi,t) \right] +D_\phi \frac{\partial^2}{\partial \phi^2}f(\mathbf{r},\phi,t) + I_\mathrm{col},
\end{align}
where $f(\mathbf{r},\phi,t)$ is the density of cells at position $\mathbf{r} = (x,y)$ and angle $\phi$ at time $t$. 
The second term in the left hand side describes the contribution where the cells keep moving with speed $v_0$ in the direction specified by the angle $\phi$ with $\mathbf{e}_\phi$ being the unit vector along the angle $\phi$.
The third term in the left hand side describes chiral motion of cells, which bends the motion of cells, with strength of the chirality characterized by the parameter $\theta$. 
The first term in the right hand side accounts for the stochastic flipping of the direction of motion with timescale $\tau$.
The second term in the right hand side is the rotational diffusion term with strength $D_\phi$.
The term $I_\mathrm{col}$ is the collision integral, for which we only keep contributions that are essential in obtaining the nematic order, as we explain later.

To obtain an effective hydrodynamic equation, we consider the Fourier expansion: 
\begin{align}
	f(\mathbf{r},\phi,t) = \frac{1}{2\pi}\sum_{k = 0}^\infty f_k(\mathbf{r},t) e^{-ik\phi}
\end{align}
and derive an equation governing the first three terms, $k = 0, 1, 2$, of the Fourier coefficients.
Note that since $f(\mathbf{r},\phi,t)$ is real, the Fourier coefficients satisfy $f_{-k} = f_k^*$.
Physically, the $k = 0$ component describes the density of particles, $\rho = f_0$. The $k = 1$ component is related to the polar vector $\mathbf{P} = (P_x, P_y)$ through $\rho \mathbf{P} = (\mathrm{Re}f_1, \mathrm{Im}f_1)$. The $k = 2$ component is related to the nematic tensor $\mathbf{Q}$ through $\rho Q_{xx} = \mathrm{Re}f_2/2$ and $\rho Q_{xy} = \mathrm{Im}f_2/2$. Expanding the Boltzmann equation in Fourier series and taking $k = 0, 1, 2$, we obtain
\begin{align}
	\partial_t f_0 &= -v_0 \partial_x \mathrm{Re}f_1 - v_0 \partial_y \mathrm{Im}f_1, \notag \\
	\partial_t f_1 &= -\frac{2}{\tau} f_1 -D_\phi f_1 - i(\theta + \nu_1 \Delta) f_1 - \frac{v_0}{2}\partial_x f_0 - i \frac{v_0}{2} \partial_y f_0 - \frac{v_0}{2}\partial_x f_2 + i \frac{v_0}{2} \partial_y f_2 	\label{hydroequations} \\ 
	\partial_t f_2 &= -4D_\phi f_2 -2i(\theta + \nu_2 \Delta) f_2 - \frac{v_0}{2}\partial_x f_1 - i\frac{v_0}{2}\partial_y f_1 + \nu \Delta f_2 + \left( \mu[\rho] - \xi |f_2|^2 \right)f_2, \notag
\end{align}
where $\Delta \equiv \partial_x^2 + \partial_y^2$ is the Laplacian, and the diffusion coefficient
\begin{align}
	\nu = \frac{v_0^2/4}{2/\tau + 9D_\phi}
	\label{diffusionconstant}
\end{align}
comes from considering the next order term, $f_3$, in the Boltzmann equation and solving for $f_3$ assuming $\partial_t f_3 = 0$ and ignoring $f_4$ and higher order terms.
In the same way, we obtain a term involving $\nu_2 \Delta$ in the equation for $f_2$. 
When defining $l=10$ $\mu$m as the minimal length scale (corresponding to the cell size) and normalizing the equation using $l$, we find that $\nu_2$ is overestimated and can become larger than $\theta$. For the calculations we perform below, we choose the sign of $\nu_2$ to be the same as $\theta$, as expected from the Boltzmann equation, but the value of normalized $\nu_2$ to be smaller than $\theta$. In the equation for $f_1$ we also have a term proportional to $\nu_1 \Delta$. This term does not come from the Boltzmann equation, but we include this term since it is necessary to regularize the theory at small length scale. A similar situation has been analyzed in more detail in~\cite{souslov_topological_2019}. It is natural to take the sign of $\nu_1$ to be the same as $\theta$, by considering the direction of the pressure that a rotating object feels inside the active fluids. These terms, $\nu_1$ and $\nu_2$, represent non-dissipative forces and torques originating from the chiral motion, where $\nu_1$ is also known as the odd (Hall) viscosity in fluid mechanics.

The final term in the equation for $f_2$ is derived from the collision integral; when the chirality is absent, this term is responsible for producing the nematic order,
\begin{align}
	f_0 &= \rho_0, & f_1 &= 0, & |f_2| &= \sqrt{\frac{\mu[\rho_0] - 4D_\phi}{\xi}} \equiv \rho_0 r.
\end{align}
For the parameters $\mu [\rho]$ and $\xi$ in the term coming from the collision integral, we use microscopic estimates from the Boltzmann-Ginzburg-Landau approach~\cite{peshkov_boltzmann-ginzburg-landau_2014,patelli2019understanding}:
\begin{align}
	\mu [\rho] &= \rho v_0 \frac{16}{3\pi}\left[ 2\sqrt{2}- \frac{12}{5}\right],
	&
	\xi &= v_0^2 \frac{128}{15\pi}\frac{\frac{16}{35\pi}\left( 6\sqrt{2} - \frac{4}{9}\right)}{\frac{16}{15\pi} \frac{176}{21}\rho_0 v_0 + 16 D_\phi}.
\end{align}

With this background, we analyze the fluctuations of $f_0$, $f_1$, and $f_2$ around the nematically ordered state, which should be achieved under the stripe boundary condition (boundary at $x = \pm L/2$). We assume that the nematic order is built along $y$-direction, so that in the absence of fluctuations, $f_2 = -\rho_0 r < 0$.
We expand up to first order in fluctuations of density, polar vector, and nematic tensor as $f_0 = \rho_0 + \delta \rho$, $f_1 = \rho_0 (\delta P_x + i \delta P_y)$, and $f_2 = -\rho_0 r + \delta f_2 = -\rho_0 r - r \delta \rho + \rho_0 (\delta Q_{xx} + i \delta Q_{xy})$.
Keeping up to the second order in these fluctuations and $\theta$ and $r$, we obtain
\begin{align}
	\partial_t
	\begin{pmatrix}
	\delta \rho/\rho_0 \\ \delta P_x \\ \delta P_y \\ \delta Q_{xx} \\ \delta Q_{xy}
	\end{pmatrix}
	=&
	\begin{pmatrix}
	0 & -v_0 \partial_x & -v_0 \partial_y & 0 & 0 \\
	-\frac{v_0}{2}(1-r)\partial_x & -\frac{2}{\tau} - D_\phi & \theta + \nu_1\Delta & -\frac{v_0}{2}\partial_x & -\frac{v_0}{2}\partial_y \\
	-\frac{v_0}{2}(1+r)\partial_y & -(\theta + \nu_1\Delta) & -\frac{2}{\tau} -D_\phi& \frac{v_0}{2}\partial_y & -\frac{v_0}{2}\partial_x \\
	\left( \mu[\rho_0] - 8 D_\phi -\nu \Delta\right)r & -\frac{v_0}{2}(1+2r)\partial_x & \frac{v_0}{2}(1-2r)\partial_y & -2(\mu[\rho_0]-4D_\phi) + \nu \Delta & 2(\theta + \nu_2\Delta) \\
	0 & -\frac{v_0}{2}\partial_y & -\frac{v_0}{2}\partial_x & -2(\theta + \nu_2\Delta) & \nu \Delta
	\end{pmatrix}
	\begin{pmatrix}
	\delta \rho/\rho_0 \\ \delta P_x \\ \delta P_y \\ \delta Q_{xx} \\ \delta Q_{xy}
	\end{pmatrix}
	\notag \\
	&+
	\begin{pmatrix}
	0 \\ 0 \\ 0 \\ 0 \\ 2\theta r
	\end{pmatrix}. \label{theequation}
\end{align}
This is the equation whose steady-state and topological properties we are going to analyze. In what follows, we take cell size 10 $\mu$m as unit of length, and one hour as the unit of time. This yields $v_0 = 4$ and $\rho_0 = 0.4$, which are the parameters adopted in the agent-based simulations and roughly corresponds to the experimental situation.
Taking $D_\phi = 0.19$, which reproduces experimental results reasonably well as we see below, the strength of the nematic order in our system is $r \approx 0.61$.

\subsection{Steady-state profile}

We first analytically explore the chiral edge flow as a steady-state solution of the linearized hydrodynamic equation~(\ref{theequation}).
We look for a solution which is time-independent and homogeneous along the $y$-direction, and also assume $\delta P_x = 0$.
Then, the linearlized hydrodynamic equations become:
\begin{align}
	&-\frac{v_0}{2}(1-r)\partial_x \frac{\delta \rho}{\rho_0} +\left( \theta + \nu_1\partial_x^2 \right)\delta P_y  -\frac{v_0}{2}\partial_x \delta Q_{xx} = 0 \label{eq:rho}\\
	&-\left(\frac{2}{\tau} + D_\phi\right)\delta P_y -\frac{v_0}{2}\partial_x \delta Q_{xy} = 0  \label{eq:py}\\
	& (mr -\nu r \partial_x^2) \frac{\delta \rho}{\rho_0} + \left( -2(\mu[\rho_0] - 4D_\phi) + \nu \partial_x^2 \right) \delta Q_{xx} + 2(\theta + \nu_2\partial_x^2) \delta Q_{xy} = 0 \label{eq:qxx}\\
	& -\frac{v_0}{2}\partial_x \delta P_y -2(\theta + \nu_2\partial_x^2) \delta Q_{xx} + \nu \partial_x^2 \delta Q_{xy} = -2\theta r, \label{eq:qxy}
\end{align}
where we set $m \equiv \mu[\rho_0] - 8 D_\phi $ for notational simplicity. 
Note that Eq.~\eqref{eq:py} represents the so-called active current (Eq.~(1) in main text).
We can find an analytical solution of the set of equations Eq.~(\ref{eq:rho})-(\ref{eq:qxy}) if we neglect $-\nu r \partial_x^2 \delta \rho$ in Eq.~(\ref{eq:qxx}).
Therefore, for the sake of analytical derivation, we ignore the term $-\nu r \partial_x^2 \delta \rho$ in Eq.~(\ref{eq:qxx}).
We have checked that the numerically obtained band structure is largely unaltered by the term $-\nu r \partial_x^2 \delta \rho$.

To solve Eqs.~(\ref{eq:rho})-(\ref{eq:qxy}), we first write $\delta P_y$ in terms of $\delta Q_{xy}$ using Eq.~(\ref{eq:py}).
Then, combining Eq.~(\ref{eq:rho}) and Eq.~(\ref{eq:qxx}), we can eliminate $\delta \rho$.
Combining the remaining equations, we can finally write a closed differential equation for $\delta Q_{xx}$:
\begin{align}
	\left( \alpha \partial_x^4 - \beta \partial_x^2 + \gamma \right) \delta Q_{xx} = \gamma r,
\end{align}
where
\begin{align}
	\alpha &= \nu \left( \nu + \frac{v_0^2}{4(2/\tau + D_\phi)}\right) + 2\nu_2 \left( 2\nu_2 - \frac{m r \nu_1}{(2/\tau + D_\phi)(1-r)} \right), \\
	\beta &= \frac{2m r}{(2/\tau + D_\phi)(1-r)}\theta (\nu_1 + \nu_2) - 8\theta \nu_2 + \left( 2(\mu [\rho_0]-4D_\phi) + \frac{mr}{1-r} \right) \left( \nu + \frac{v_0^2}{4(2/\tau + D_\phi))}\right), \\
	\gamma &= 2\theta^2 \left( 2 - \frac{m r}{(2/\tau + D_\phi)(1-r)}\right).
\end{align}
A general solution of this differential equation can be written as
\begin{align}
	\delta Q_{xx} = r + C_1 e^{\lambda_+ x} + C_2 e^{\lambda_- x} + C_3 e^{-\lambda_+ x} + C_4 e^{-\lambda_- x},
\end{align}
where 
\begin{align}
	\lambda_\pm \equiv \sqrt{\frac{\beta \pm \sqrt{\beta^2 - 4\alpha \gamma}}{2\alpha}},
\end{align}
and $C_1$, $C_2$, $C_3$, and $C_4$ are constants of integral to be determined from boundary conditions.

For boundary conditions, we assume $\delta Q_{xx} = \delta Q_{xy} = 0$ at the boundaries $x = \pm L/2$,. We further assume that the integral of the density fluctuation is zero, that is, the number of cells is conserved: $\int_{-L/2}^{L/2} dx \delta \rho = 0$. With these boundary conditions, we can fully determine the constants of integral. The steady-state solution with these boundary conditions is
\begin{align}
	\frac{\delta \rho}{\rho_0} &= \frac{\rho_c}{\rho_0}
	+ \frac{2 C_1}{mr}\left( 2(\mu[\rho_0] - 4D_\phi) - \nu \lambda_+^2 - \frac{4}{\lambda_+^2}\frac{(\theta + \nu_2  \lambda_+^2)^2}{\nu + \frac{v_0^2}{4(2/\tau + D_\phi)}} \right) \cosh (\lambda_+ x)
	\notag \\
	&\hspace{1cm}+ \frac{2 C_2}{mr}\left( 2(\mu[\rho_0] - 4D_\phi) - \nu \lambda_-^2 - \frac{4}{\lambda_-^2} \frac{(\theta + \nu_2  \lambda_-^2)^2}{\nu + \frac{v_0^2}{4(2/\tau + D_\phi)}} \right) \cosh (\lambda_- x) ,\\
	\delta P_y &= -\frac{2v_0}{(2/\tau + D_\phi)\nu + v_0^2/4}\left( \frac{C_1}{\lambda_+}(\theta + \nu_2 \lambda_+^2) \sinh (\lambda_+ x) + \frac{C_2}{\lambda_-}(\theta + \nu_2 \lambda_-^2) \sinh (\lambda_- x) \right) ,\\
	\delta Q_{xx} &= r + 2 C_1 \cosh (\lambda_+ x) + 2 C_2 \cosh (\lambda_- x) , \\
	\delta Q_{xy} &= \frac{2(\mu[\rho_0] - 4D_\phi) r - mr\rho_c/\rho_0}{2\theta} + 4 C_1 \frac{\theta + \nu_2 \lambda_+^2}{(\nu + \frac{v_0^2}{4(2/\tau + D_\phi)})\lambda_+^2}\cosh (\lambda_+ x) + 4 C_2 \frac{\theta + \nu_2 \lambda_-^2}{(\nu + \frac{v_0^2}{4(2/\tau + D_\phi)})\lambda_-^2}\cosh (\lambda_- x) ,
\end{align}
where the coefficients $C_1$ and $C_2$ are
\begin{align}
	C_1 &= \frac{1}{\cosh (L\lambda_+/2)}\frac{1}{4\theta^2 (1 - \lambda_-^2/\lambda_+^2)}\left[ \left( \nu + \frac{v_0^2}{4(2/\tau + D_\phi)}\right)\left( (\mu[\rho_0] - 4D_\phi) r-\frac{mr}{2}\frac{\rho_c}{\rho_0}\right)\lambda_-^2 - (\theta + \nu_2 \lambda_-^2)2\theta r \right], \\
	C_2 &= \frac{1}{\cosh (L\lambda_-/2)}\frac{1}{4\theta^2 (1 - \lambda_+^2/\lambda_-^2)}\left[ \left( \nu + \frac{v_0^2}{4(2/\tau + D_\phi)}\right)\left( (\mu[\rho_0]- 4D_\phi) r -\frac{mr}{2}\frac{\rho_c}{\rho_0}\right)\lambda_+^2 - (\theta + \nu_2 \lambda_+^2)2\theta r \right],
\end{align}
and $\rho_c$ is a constant to be determined from the condition $\int_{-L/2}^{L/2} dx \delta \rho = 0$.

In Extended Data Fig.~4, we plot the steady-state for $\theta = 0.2$, $\nu_1 = \nu_2 = 0.02$ with different $L$. We see that  $\delta P_y$, which is the average velocity of the steady-state in $y$-direction, as well as the density fluctuation $\delta \rho/\rho_0$ is localized at both edges. The steady-state has velocity in positive $y$-direction at the right edge, and negative direction at the left edge, which is consistent with what we observe in experiment and numerical simulation. 

From the steady-state results, we obtain $\delta \rho/\rho_0 \sim L^2 \theta^2/m \nu$, $\delta P_y \sim L \theta^2 / v_0$, and $\delta Q_{xx} \sim \delta Q_{xy} \sim L^2 \theta / \nu$ in the lowest order of $\theta$. This corresponds to the fact that the linearization around the nematically ordered state with no flow is justified only when $\theta$ is smaller than the inverse of the time scales in the system. When fixing $\theta$ and increasing the width of the system $L$, the condition for the linearization becomes violated, and the nonlinear terms in the hydrodynamic equations should be taken into account. Nevertheless, we here continue using the linear equation for the sake of simplicity, and with the expectation that the basic physics will be unaltered by the nonlinear terms since the steady-state solutions are still qualitatively similar to the experimental and numerical observations. We note that the large $L$ limit of the linear equation is still well-behaved since $\delta \rho/\rho_0$, $\delta P_y$, $\delta Q_{xx}$, and $\delta Q_{xy}$ all converge to $O(1)$ values.

\subsection{Edge-localized topological Kelvin mode}

The steady-state obtained above is the time-independent solution of the linearized hydrodynamic equation. 
We now show that there are also normal modes of the equation without a source term, which is localized at the edges and propagates unidirectionally. These modes turn out to have a topological origin, and is essentially the Kelvin modes known in geophysics.

As before, we consider a configuration where the system is bounded in the $x$-direction but long in the $y$-direction.
The momentum $k_y$ along $y$-direction is thus a good ``quantum" number, and we can replace $\partial_y$ by $ik_y$ in the equation. 
We look for the solution with $ \delta P_x=0$, and also neglect, for simplicity, the mixing effects between the momentum and nematic variables. We later compare the full numerical solution of the model with the analytical solution obtained here and confirm that the analytical solution matches the full numerical solution well, validating the approximation we employ here.

The simplified linearized hydrodynamic equation then takes the following form:
\begin{align}
-i E(k_y)
\begin{pmatrix} \delta \rho/\rho_0 \\ 0 \\ \delta P_y \end{pmatrix}
=
\begin{pmatrix}
	0 & -v_0 \partial_x & -iv_0 k_y \\
	-\frac{v_0}{2}(1-r)\partial_x & -\frac{2}{\tau}-D_\phi &  \theta +\nu_1 (\partial_x^2 - k_y^2) \\
	-i\frac{v_0}{2}(1+r)k_y & -\left[ \theta + \nu_1 (\partial_x^2 - k_y^2)\right] & -\frac{2}{\tau} - D_\phi
\end{pmatrix}
\begin{pmatrix} \delta \rho/\rho_0 \\0 \\  \delta P_y  \end{pmatrix}.
\end{align}
By solving the linearized equations, we obtain the dispersion relation:
\begin{align}
	E_\pm (k_y) = -\frac{i}{2}\left( \frac{2}{\tau} + D_\phi \right) \pm \frac{1}{2}\sqrt{2(1+r)v_0^2 k_y^2 - \left( \frac{2}{\tau}+D_\phi\right)^2}.
\end{align}
We find that the dispersion relation does not depend on $\theta$, and becomes flat at $k_y \approx 0$, i.e., ${\rm Re}E(k_y)=0$ if $|k_y|$ is small.
In this momentum range, the waves are overdamped.
On the other hand, outside of the range where ${\rm Re}E(k_y) = 0$, waves propagate with finite life-time.

Next we discuss the localization of the waves.
Since we are interested in the propagating wave, we focus on the momentum range where $|k_y|$ is sufficiently large such that the dispersion relation becomes
\begin{align}
E_\pm (k_y)\sim \pm \sqrt{\frac{1+r}{2}} v_0 k_y. \label{kelvindispersion}
\end{align}
The amplitude of the waves satisfies the differential equation
\be
0 =\left(\nu_1\partial_x^2-\frac{v_0^2 (1-r) k_y}{2 E_\pm(k_y)}\partial_x+\theta-\nu_1k_y^2\right) \delta P_y .
\ee
Assuming that $\nu_1$ is small and keeping leading orders in $\nu_1$, we obtain
\begin{align}
	\delta P_y = D_1 \exp \left\{ \left[ \frac{(1-r)v_0^2 k_y}{2 E_\pm (k_y)}\frac{1}{\nu_1} - \frac{2E_\pm (k_y)}{(1-r)v_0^2 k_y} \theta \right]x \right\} + D_2 \exp \left[ \frac{2E_\pm (k_y)}{(1-r)v_0^2 k_y}\theta x \right],
\end{align}
where $D_1$ and $D_2$ are constants to be determined from boundary conditions.

From this expression of $\delta P_y$, we observe that if the sign of $\theta$ and $\nu_1$ are the same, the up-going wave [described by $E_+ (k_y)$] and the down-going wave [described by $E_- (k_y)$] are localized to the opposite sides of the edges, so that unidirectional propagating waves appears at the edges.
As we see below, this is nothing but the topological edge modes predicted from the bulk-edge-correspondence, and is what we observed in power spectrum obtained from experiments and numerical simulations (Fig.~4, Extended Data Figs.~5,6,7).

A noticeable feature we can observe here is that the dispersion relation itself does not depend on the strength of the chirality $\theta$, but the localization length does. 

\subsection{Effective Hamiltonian and its topological properties}

We now discuss the numerical band structure and topological properties of the linearized hydrodynamic equation.
In order to make connection with quantum mechanics and the physics of topological insulators, we multiply both sides of the equation by an imaginary unit $i$.
Defining 
\begin{align}
	\Psi &\equiv \begin{pmatrix} \delta \rho/\rho_0 \\  \delta P_x \\  \delta P_y \\  \delta Q_{xx} \\  \delta Q_{xy} \end{pmatrix},
	&
	\mathrm{and}& &
	s &\equiv \begin{pmatrix} 0 \\ 0 \\ 0 \\ 0 \\ i 2 \theta r \end{pmatrix},
\end{align} 
the equation takes the following form:
\begin{align}
	i\partial_t \Psi = \mathcal{H} \Psi + s, \label{schrodinger}
\end{align}
where 
\begin{align}
	\mathcal{H} \equiv i\mathcal{M} 
	=
	\begin{pmatrix}
	0 & v_0 \hat{p}_x & v_0 \hat{p}_y & 0 & 0 \\
	\frac{v_0}{2}(1-r)\hat{p}_x & -\left( \frac{2}{\tau} + D_\phi \right)i & i\left[ \theta - \nu_1 \hat{p}^2\right] & \frac{v_0}{2}\hat{p}_x & \frac{v_0}{2}\hat{p}_y \\
	\frac{v_0}{2}(1+r)\hat{p}_y & -i\left[ \theta - \nu_1 \hat{p}^2\right] & -\left( \frac{2}{\tau} + D_\phi \right)i & -\frac{v_0}{2}\hat{p}_y & \frac{v_0}{2}\hat{p}_x \\
	i \left[ m  + \nu \hat{p}^2 \right]r & \frac{v_0}{2}(1+2r)\hat{p}_x & -\frac{v_0}{2}(1-2r)\hat{p}_y & -i2(\mu[\rho_0] - 4D_\phi) - i \nu \hat{p}^2 & 2i \left[ \theta - \nu_2 \hat{p}^2\right] \\
	0 & \frac{v_0}{2}\hat{p}_y & \frac{v_0}{2}\hat{p}_x & -2i \left[ \theta - \nu_2 \hat{p}^2\right] & -i \nu \hat{p}^2
	\end{pmatrix}.
\end{align}
with $\hat{p}_i \equiv -i\partial_i$ being the momentum operator and $\hat{p}^2 \equiv \hat{p}_x^2 + \hat{p}_y^2$.
The equation takes the same form as the Scrh\"odinger equation under an external source $s$, with $\Psi$ serving as a wavefunction and $\mathcal{H}$ as the Hamiltonian. Note that the matrix $\mathcal{H}$ is non-Hermitian. 

\subsubsection{Case of gapped Hamiltonian}

We first explore the band structure of the non-Hermitian Hamiltonian $\mathcal{H}$ when the strength of the chirality is large and an energy gap opens. 
We take the $y$-direction to be long so that the momentum along the $y$-direction is a good quantum number.
In the $x$-direction, we may take either a periodic boundary condition or an open boundary condition with edges. 
In Extended Data Fig.~5, we plot the energy spectrum as a function of the momentum $k_y$ along the $y$-direction, for the case of $\theta = 2$, $\nu_1 = \nu_2 = 0.2$, and $\nu = 0.5$ .
Extended Data Figs.~5a,b are the band structures when the $x$-direction is taken to have a periodic boundary condition, and Extended Data Figs.5~c,d show the band structure when $x$ is bounded by sharp edges (open boundary condition) at $x = \pm L/2$ with $L=30$.

We observe that in the presence of edges, there are modes crossing the gaps between the bands, which are the topological edge modes.
In fact, the Chern number of the lowest two bands of our non-Hermitian model is six, which agrees with the six edge states present in Extended Data Fig.~5c according to the bulk-boundary correspondence.
In Extended Data Fig.~5c, we also plot the analytical result for the edge-localized mode obtained in Eq.~(\ref{kelvindispersion}) as the red lines.
We see that one of the six topological edge modes agree well with the analytical line, confirming that this edge state is the topological Kelvin mode.
Another mode, which is dispatched from $\mathrm{Re}(E)\tau \approx \pm 6$ and extends far in momentum, corresponds to the Yanai mode also observed in geophysics~\cite{delplace_topological_2017}.
There are four additional topological edge states that are only present when the nematic tensor is included in the equation.

To see how the modes are localized at the edges, we plotted the spectrum with the colours indicating the localization at the edges (Extended Data Fig.~5e,f).
We see that all six topological edge modes with positive group velocity are localized at the right edge, whereas those with negative group are localized at the left edge. This observation is again consistent with the bulk-edge correspondence.

Topology of non-Hermitian systems has been recently attracting considerable interest~\cite{shen2018topological, gong2018topological, kawabata2019symmetry}.
In our system, we can continuously transform our non-Hermitian Hamiltonian into a Hermitian Hamiltonian without closing the gap in the real part of the complex energy spectrum. 
In the terminology of~\cite{kawabata2019symmetry}, our system is characterized by the symmetry class A with a line gap at $\mathrm{Re}E(\mathbf{k}) \approx 0$, where $E(\mathbf{k})$ is the complex energy as a function of the momentum $\mathbf{k}$.
The Hermitian Hamiltonian to which we can continuously deform our non-Hermitian Hamiltonian is
\begin{align}
	\mathcal{H}_\mathrm{Herm} = 
	\begin{pmatrix}
	0 & v_0 \hat{p}_x & v_0 \hat{p}_y & 0 & 0 \\
	v_0 \hat{p}_x & 0 & i\left[ \theta - \nu_1 (\hat{p}_x^2 + \hat{p}_y^2)\right] & \frac{v_0}{2}\hat{p}_x & \frac{v_0}{2}\hat{p}_y \\
	v_0 \hat{p}_y & -i\left[ \theta - \nu_1 (\hat{p}_x^2 + \hat{p}_y^2)\right] & 0& -\frac{v_0}{2}\hat{p}_y & \frac{v_0}{2}\hat{p}_x \\
	0 & \frac{v_0}{2}\hat{p}_x & -\frac{v_0}{2}\hat{p}_y & 0 & 2i \left[ \theta - \nu_2 (\hat{p}_x^2 + \hat{p}_y^2)\right] \\
	0 & \frac{v_0}{2}\hat{p}_y & \frac{v_0}{2}\hat{p}_x & -2i \left[ \theta - \nu_2 (\hat{p}_x^2 + \hat{p}_y^2)\right] & 0
	\end{pmatrix}.
\end{align}
In Extended Data Fig.~5g, we plot the real part of the energy spectrum as a function of a parameter which connects $\mathcal{H}$ and $\mathcal{H}_\mathrm{Herm}$; the left end of the figure corresponds to $\mathcal{H}$ and the right end to $\mathcal{H}_\mathrm{Herm}$.
We observe that the energy gap does not close when connecting $\mathcal{H}$ and $\mathcal{H}_\mathrm{Herm}$.

We note the essential difference between conventional topological lattice models and our system.
Our system, unlike systems defined on a lattice, does not have a compact momentum space.
Typically, topological invariant such as the Chern number should be defined on a compact parameter space.
The Chern number can be calculated by integrating the Berry curvature over the compact parameter space, and the integral is guaranteed to be an integer.
In our system, the momentum space is not compact, so the Chern number, calculated analogously to lattice systems, is not guaranteed to take an integer value. 
However, it has been shown that when odd viscosity terms ($\nu_1$ and $\nu_2$) are present, the Chern number is guaranteed to be an integer~\cite{souslov_topological_2019, banerjee_odd_2017,tauber_bulk-interface_2019}. 
Indeed, when we calculate the Chern number of our Hamiltonian for the energy gap around the zero energy, we obtain a value which approaches $6$ as we take into account larger areas in momentum space for integrating the Berry curvature (Extended Data Fig.~5h). We also note that when $\theta > 0$ and $\nu_2 < 0$, the Chern number of the lowest two bands are two, and there are two topological edge states. When both $\nu_1$ and $\nu_2$ are negative, there is no topological edge state and the Chern number is zero.

As plotted in Extended Data Fig.~4b, the velocity field in the $y$-direction is localized at the edges.
The integrated velocity along the $y$-direction essentially picks up the $k_y = 0$ component of the eigenstate.
Our model, at $k_y = 0$, does not have any topological edge state as one can observe from the band structure.
Therefore, we consider that the edge localization of the steady-state itself is a distinct mechanism from the topological edge states analyzed above.

In fact, we notice that many of the modes in the bands in the bulk localize at the edge in the case of the open boundary condition (Extended Data Fig.~5e). This phenomena of the bulk modes localizing in the presence of the boundary is called the non-Hermitian skin effect, and has also been recently found to have a topological explanation~\cite{Yao2018,Okuma2020}.

\subsubsection{Case where the gaps are closed}

By comparing the experiment and numerical results of the agent-based model, we estimate that the realistic value of chirality is around $\theta = 0.2$.
In Extended Data Fig.~6a, we plot the band structure for $\theta = 0.2$ and $\nu_1 = \nu_2 = 0.02$.
For $\nu$, we used the value calculated from Eq.~(\ref{diffusionconstant}).

We find that the bulk energy gap is closed, which is due to the large dissipation in the hydrodynamic equation. 
However, the mode described by Eq.~(\ref{kelvindispersion}) still exists in the spectrum (Extended Data Fig.~6c), indicating that the Kelvin mode may be surviving even in the case where the gap is closed.

To confirm this, we plotted the spectrum with the colours indicating the localization at the edges and the darkness indicating the contribution to the density fluctuation for each mode (Fig.~4a in the main text).
Here, the contribution to the density fluctuation was calculated by normalizing the wave function and looking at the $\delta \rho/\rho_0$ component.
We observe that, among all the modes which exist around zero energy, only the modes forming the large X in the figure survives after extracting modes with significant density fluctuation, which fits with the Kelvin mode [Eq.~(\ref{kelvindispersion})].
When lowering the chirality ($\theta=0.02$, Extended Data Fig.~7a), we found that the similar modes survived but in a non-localized manner.
This is consistent with the experiment, where the slope of the dispersion of the power spectrum does not seem to be altered by the change of $\theta$ by the application of Jasplakinolide except for the flip in the direction (Fig.~4e in the main text).
The existence of the wave modes for the case of small $\theta$, and the change in the localization by increasing $\theta$, were also observed in the spectral analysis of the agent-based model (Extended Data Fig.~7b,c).

\newpage

\setcounter{figure}{0}
\renewcommand{\figurename}{{\bf Extended Data Fig.}}
\renewcommand\thefigure{{\bf \arabic{figure}}}

\begin{figure*}[!t]
 \begin{center}
  \includegraphics[width=183mm]{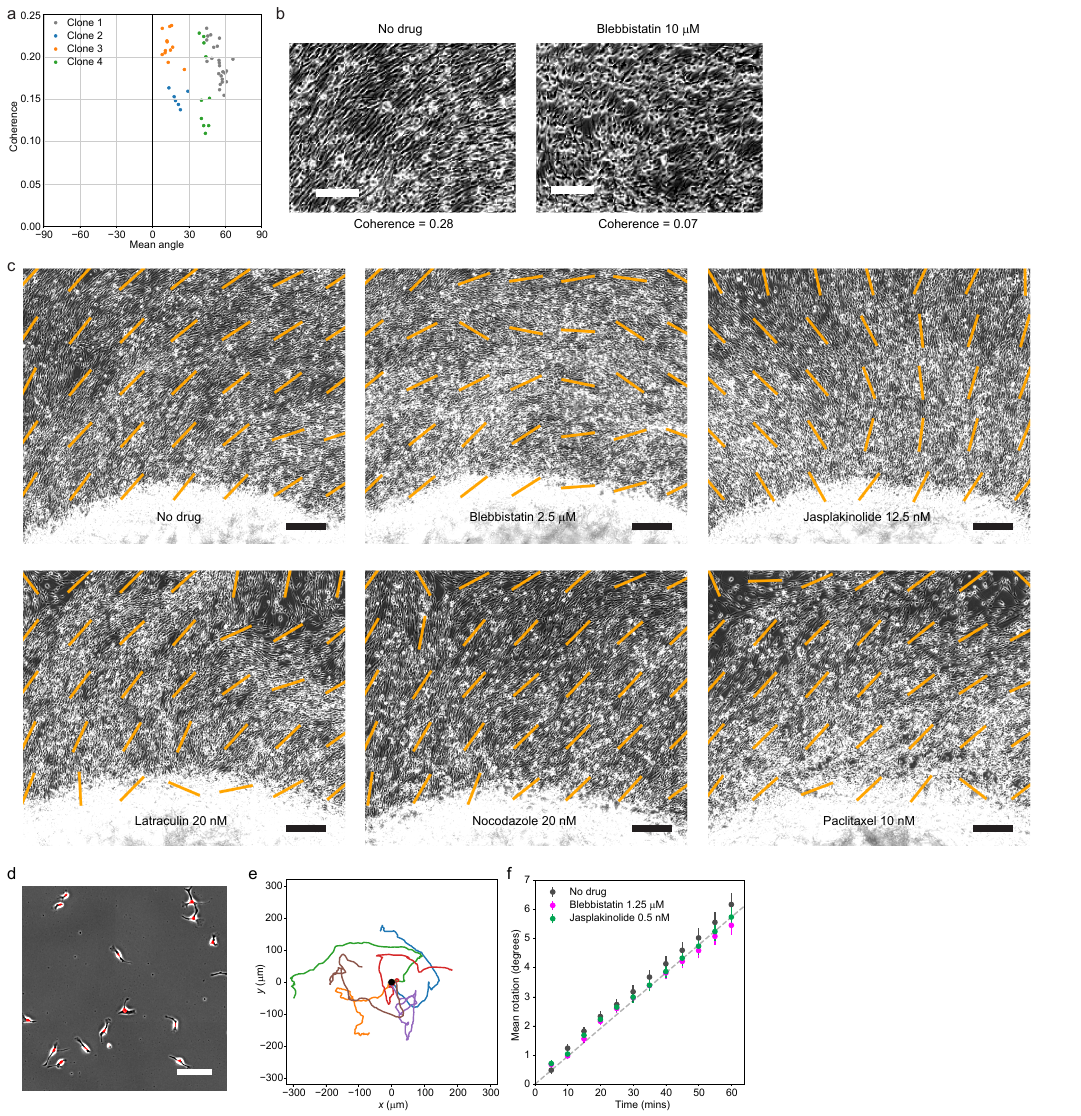}
 \end{center}
  \caption{\label{FigS1} {\bf Chirality of neural progenitor cells.}
{\bf a,} Chirality and coherence calculated from the gel experiment data using different clones of NPCs. Each dot represents a single gel drop experiment.
{\bf b,} Phase contrast images of regions with high and low coherence. The coherence was calculated by taking the spatial average of $C({\bf r})$ (Eq.~(9) in the Methods section) within the shown image region. Scale bars: 100~$\mu$m. 
{\bf c,} Phase contrast images of the gel experiment overlaid with the direction of alignment. Scale bars: 200~$\mu$m. 
{\bf d,} Example image of a sparse culture condition. Grey: phase contrast. Red: H2B-mCherry signal. Scale bar: 100~$\mu$m. 
{\bf e,} 10 hour trajectories of randomly chosen NPCs in the sparse culture condition. 
{\bf f,} Average rotation of the motion of in the NPCs in the clockwise direction. 1103 (No drug), 1085 (Blebbistatin), and  843 (Jasplakinolide) cells were used in calculating the average. The grey dashed line corresponds to 0.1 rad/hour. Error bars: s.e.m. 
}
\end{figure*}

\begin{figure}[!t]
 \begin{center}
  \includegraphics[width=89mm]{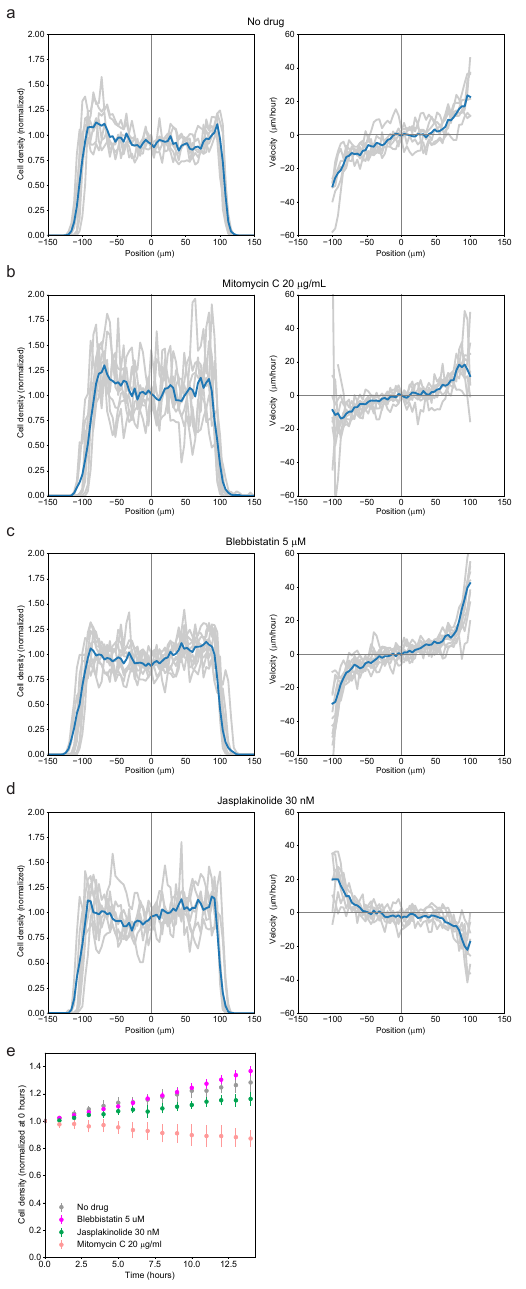}
 \end{center}
  \caption{\label{FigS2} {\bf Results of drug perturbation.}
Cell density and velocity for the stripe condition with width $L=$ 200~$\mu$m without drug ({\bf a}), with Mitomycin C 20 $\mu$g/ml ({\bf b}), Blebbistatin 5 $\mu$M ({\bf c}), and Jasplakinolide 30 nM ({\bf d}).
Grey lines correspond to data from single stamp regions, and the blue line is the average over the regions.
{\bf e,} Time course of the mean cell density upon application of drugs.
}
\end{figure}

\begin{figure*}[!hbt]
 \begin{center}
  \includegraphics[width=183mm]{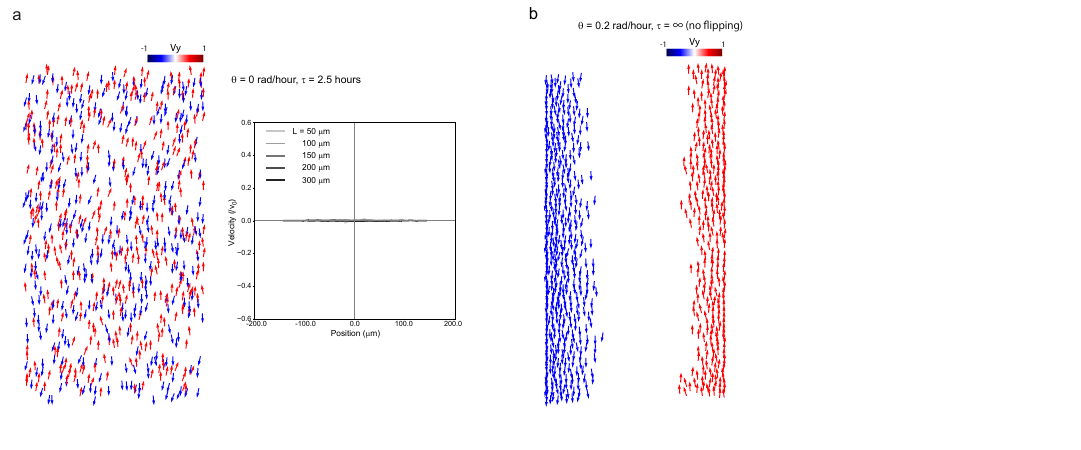}
 \end{center}
  \caption{\label{FigS3} {\bf Simulation results of the agent-based model for different parameters of chirality and flipping rate.}
{\bf a,} Configuration (left) for the case of $\theta =0$ rad/hour and $\tau =2.5$ hours. The particles show clear nematic order but no net flow (right).
{\bf b,} Configuration for the case of $\theta = 0.2$ rad/hour and $\tau = \infty$ (i.e., no stochastic flipping). The particles form unidirectionally transporting lanes near the edges.
}
\end{figure*}

\begin{figure*}[!hbt]
 \begin{center}
  \includegraphics[width=183mm]{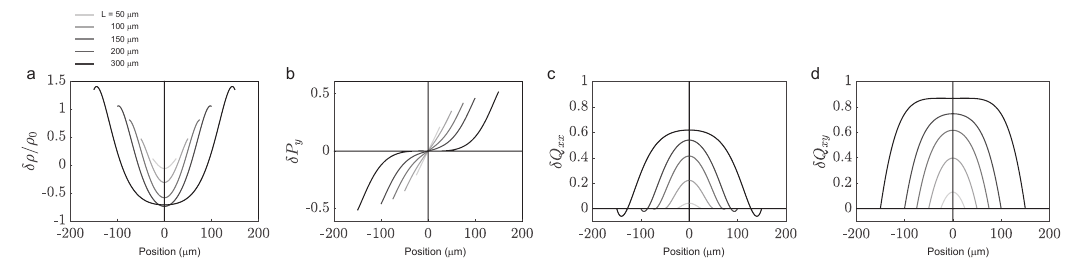}
 \end{center}
  \caption{\label{FigS4} {\bf Steady state profile of the linearized hydrodynamic equation.}
  {\bf a-d,} The density $\delta \rho / \rho_0$ ({\bf a}), flow velocity in the $y$-direction (normalized by the cell velocity $v_0$) $\delta P_y$ ({\bf b}), nematic order parameters $\delta Q_{xx}$ ({\bf c}), and $\delta Q_{xy}$ ({\bf d}) plotted against the $x$-position for $L=50, 100, 150, 200, 300$ $\mu$ m. We used the results described in Sec.IIIA of the Supplemental Material, with the parameters $\theta = 0.2$ rad/hour, $\nu_1 = \nu_2 = 2$ rad$\cdot$ $\mu$m$^2$/hour.}
\end{figure*}

\begin{figure*}[!hbt]
 \begin{center}
  \includegraphics[width=183mm]{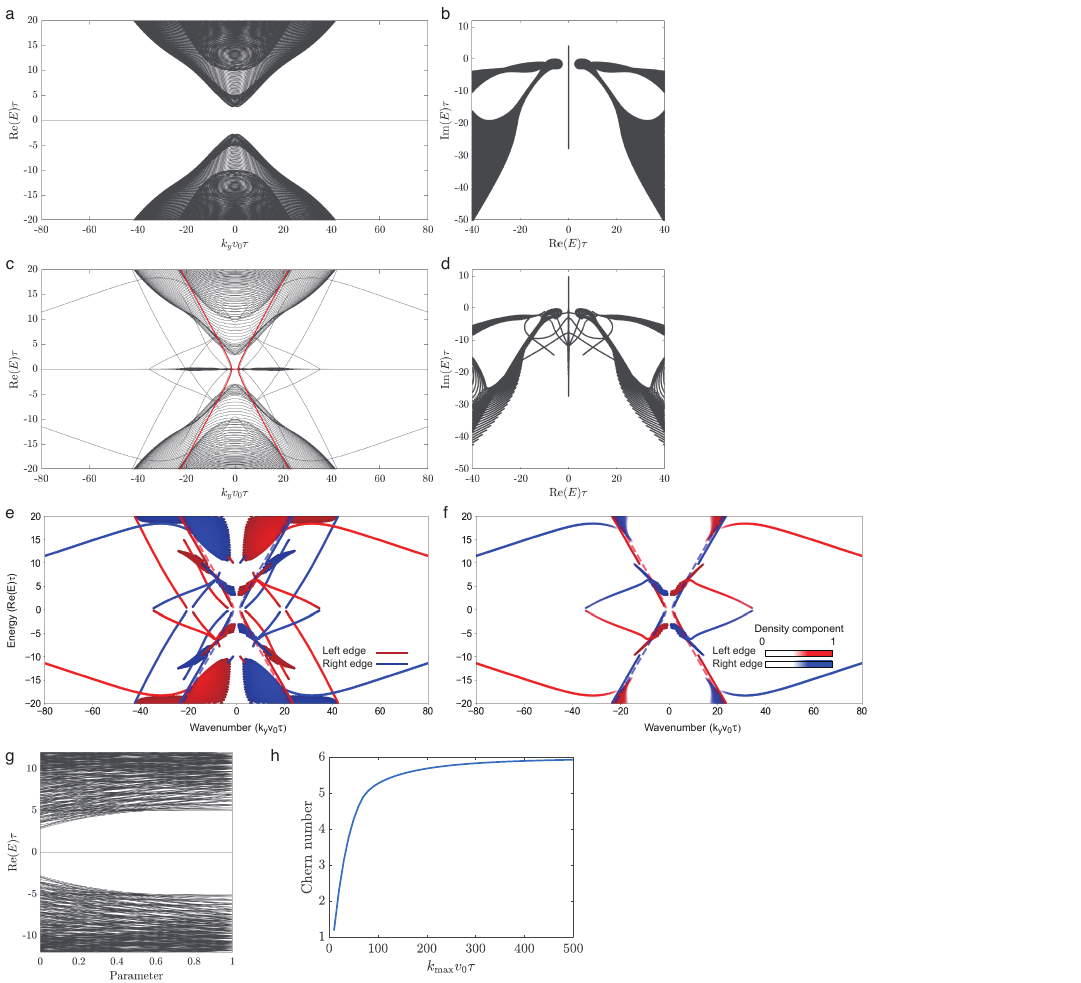}
 \end{center}
  \caption{\label{FigS5} {\bf Band structures of the effective non-Hermitian Hamiltonian for large chirality.}
{\bf a,} The band structure under periodic boundary condition as a function of the wavenumber $k_y$ along the $y$-direction showing the real part of the complex energy, calculated for the non-Hermitian Hamiltonian in the case of $\theta = 2.0$ (2.0 rad/hour), $\nu_1 = \nu_2 = 0.2$ (20 rad$\cdot$ $\mu$m$^2$/hour), and $\nu =0.5$ (50 $\mu$m$^2$/hour).
{\bf b,} The energy plotted in the complex plane for the same setup as ({\bf a}).
{\bf c,} Same setup and plot as {\bf a} but calculated under the stripe (open) boundary condition with width $L = 300$ $\mu$m.
The solid red lines correspond to the analytical result for edge-localized mode $E_+ (k_y)$ obtained in as Eq.~(3) in the main text.
{\bf d,} The energy plotted in the complex plane for the same setup as ({\bf c}).
{\bf e,} Same setup and plot as {\bf c} but colour coded according to the localization at the edge. 
{\bf f,} Same setup and plot as {\bf c} but colour coded according to the localization at the edge (colour) and the contribution of the modes to the density fluctuation (intensity).
{\bf g,} Real part of the energy calculated under periodic boundary condition, demonstrating that the non-Hermitian Hamiltonian can be adiabatically transformed to a Hermitian Hamiltonian (Eq.~(S32) in Supplementary Text) without closing the energy gap. The left end of the figure corresponds to the non-Hermitian Hamiltonian, and the right end corresponds to the Hermitian counterpart, connected by a single parameter.
{\bf h,} Chern number of the non-Hermitian Hamiltonian calculated with varying maximum wavenumber ($k_{\rm max}$) in the integral.
}
\end{figure*}

\begin{figure*}[!hbt]
 \begin{center}
  \includegraphics[width=183mm]{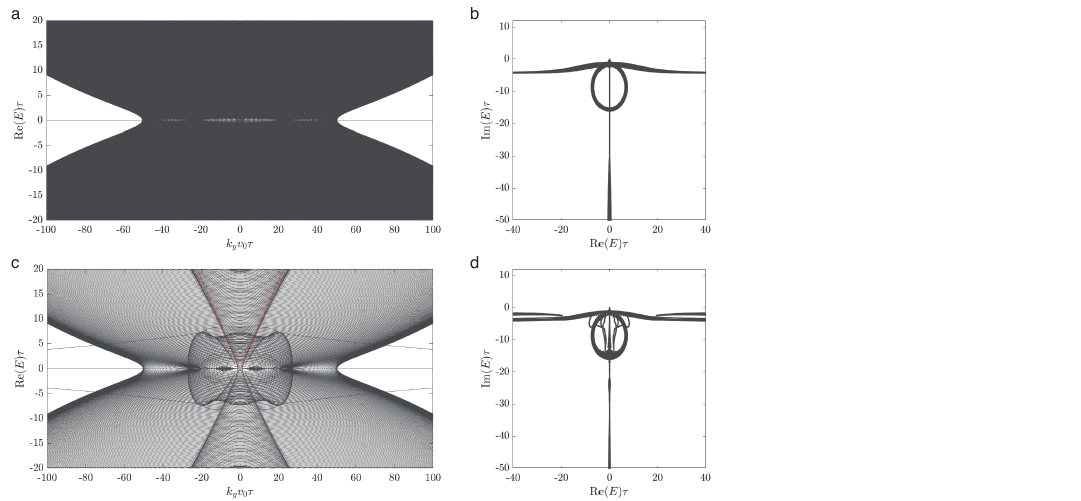}
 \end{center}
  \caption{\label{FigS6} {\bf Band structures of the effective non-Hermitian Hamiltonian for an experimentally relevant value of chirality}.
{\bf a,} The band structure under periodic boundary condition as a function of the wavenumber $k_y$ along the $y$-direction showing the real part of the complex energy, calculated for the non-Hermitian Hamiltonian in the case of $\theta = 0.2$ (0.2 rad/hour), $\nu_1 = \nu_2 = 0.02$ (2 rad$\cdot$ $\mu$m$^2$/hour).
{\bf b,} The energy plotted in the complex plane for the same setup.
{\bf c,} Same setup and plot as {\bf a} but calculated under the stripe (open) boundary condition with width $L = 300$ $\mu$m.
The solid red lines correspond to the analytical result for edge-localized mode $E_+ (k_y)$ obtained in as Eq.~(3) in the main text.
{\bf d,} The energy plotted in the complex plane for the same setup as {\bf c}.
}
\end{figure*}

\begin{figure*}[!hbt]
 \begin{center}
  \includegraphics[width=183mm]{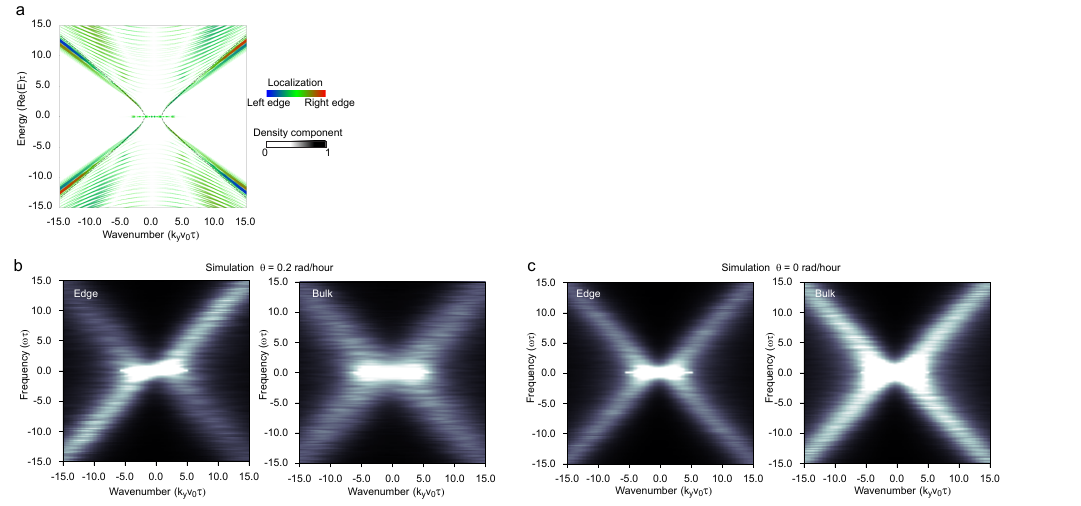}
 \end{center}
  \caption{\label{FigS7} {\bf Band structure and localization.}
{\bf a,} Energy spectrum of the effective non-Hermitian Hamiltonian calculated with the stripe boundary condition with $\theta=0.02$ rad/hour. Colour indicates the position of the localization of the modes, and the intensity of the colour indicates the contribution of the mode to the density fluctuation.
{\bf b,} Power spectrum of the agent density calculated for the edge and bulk regions from the numerical simulation with $\theta=0.2$ rad/hour.
{\bf c,} Power spectrum of the agent density calculated for the edge and bulk regions from the numerical simulation with $\theta=0$ rad/hour.
}
\end{figure*}

\end{document}